\documentclass[twocolumn]{aa} 

\usepackage{graphicx} 
\usepackage{caption}
\usepackage{subcaption}
\usepackage{float}
\usepackage{multirow}
\usepackage{array, multirow, tabularx}
\usepackage{adjustbox} 
\usepackage{amsmath}
\usepackage[normalem]{ulem}
\usepackage{cancel}
\usepackage{txfonts}
\usepackage[hidelinks]{hyperref}
\hypersetup{
    colorlinks,
    citecolor=blue,
    filecolor=blue,
    linkcolor=blue,
    urlcolor=blue
}
\begin{document} 
\pagenumbering{arabic}

   \title{Probing millisecond magnetar formation in binary neutron star mergers through X-ray follow-up of gravitational wave alerts}

   \author{C. Plasse\thanks{corresponding author: clara.plasse@outlook.fr}
          \inst{1} \and A. Reboul-Salze \inst{2} \and J. Guilet \inst{1} \and D. Götz \inst{1} \and N. Leroy \inst{3} \and  R. Raynaud \inst{1} \and M. Bugli \inst{4,5} \and  T. Dal Canton \inst{3}}

  \institute{Université Paris-Saclay, Université Paris Cité, CEA, CNRS, AIM, 91191, Gif-sur-Yvette, France \and
     Max Planck Institute for Gravitational Physics (Albert Einstein Institute), D-14476 Potsdam, Germany \and
   Université Paris-Saclay, CNRS/IN2P3, IJCLab, 91405 Orsay, France \and
   Institut d’Astrophysique de Paris, Sorbonne Université, CNRS, UMR 7095, 98 bis bd Arago, 75014 Paris, France \and
   Dipartimento di Fisica, Università di Torino, Via P. Giuria 1, I-10125 Torino, Italy
   }

   \date{Received January 2026; accepted xxx}

\titlerunning{Probing millisecond magnetar formation in BNS merger through X-ray follow-up of GW alerts}

  \abstract
    {The nature of the remnant of a binary neutron star (BNS) merger is uncertain. Although it certainly is a black hole in the cases of the most massive BNSs, X-ray light-curves from short gamma-ray burst afterglows suggest a neutron star (NS) as a viable candidate for the merger remnant and central engine of these transients. 
    When jointly observed with gravitational waves (GWs), X-ray light-curves from BNS merger events could provide critical constraints on the remnant nature.
    }
    {We assess the current and future capabilities for detecting an NS remnant through X-ray observations following GW detections.
    }
    {To this end, we simulated GW signals from BNS mergers and the subsequent X-ray emission from newborn millisecond magnetars. We modeled the GW detectability for the current and next-generation GW interferometers, and we reproduced the X-ray emission using a dedicated numerical code that models magnetar spin-down and ejecta dynamics informed by numerical relativity simulations.
    }
   {In our simulations, 2\% $-$ 16\% of the BNS mergers form millisecond magnetars. Up to $\sim$ 70\% of these might be detectable, which means up to 1.0$^{+0.3}_{-0.3}$ millisecond magnetar detections per year with instruments such as SVOM/MXT during the LIGO Virgo KAGRA LIGO India (LVKI) O5 run. The best detectability occurs about two hours post merger. For next-generation GW interferometers, this rate might increase by up to three orders of magnitude, with the peak detectability three to four hours post merger. We also explored how the magnetar magnetic field strength and observer viewing angle affect detectability, and we discuss optimized observational strategies.
   }
   {Although more likely with upcoming GW interferometers, the detection of the spin-down emission of a millisecond magnetar may already be within reach. This warrants sustained theoretical and observational efforts given the profound implications for mergers, gamma-ray bursts, and NS physics of a single detection.
   }


   \maketitle
%

\section{Introduction}
    Short gamma-ray bursts (sGRBs) are commonly associated with the merger of compact objects, namely two neutron stars (NSs; \citealt{grb_classification, shortvslong_sumup}), as confirmed with by the event on August 17, 2017 \citep{multimessagers, 170817, 170817_kn}. The merger is followed by a collimated relativistic ejection along the rotation axis of the remnant, which is detected as prompt emission for an observer on the axis of this jet \citep{magnetar_central_engine_swift, Metzger_review}. The fainter and rapidly fading X-ray afterglow follows the prompt emission. 

    Most often proposed to be a black hole (BH), the binary neutron star (BNS) merger remnant might also be an NS (stable or unstable). In particular, the possibility that the merger might produce a rapidly spinning high magnetic field NS (i.e., a millisecond magnetar) has been suggested by \cite{Zhang_2006} to interpret some of the X-ray features in sGRB afterglows. Concretely, an X-ray afterglow light-curve comprising a plateau followed by a steep decay is more consistently described by a millisecond magnetar that is long-lived and unstable than a super-accreting BH \citep{Zhang_2006, rowlinson_2013}. A millisecond magnetar is expected to store a considerable amount of energy in the form of rotational energy, up to 10$^{53}$~erg \citep{10_53_erg}. 
    The extraction of the millisecond magnetar rotational energy by the dipolar component of the magnetic field can power the plateau light-curve feature. If this plateau is followed by a steep luminosity decay, it is naturally explained by the collapse of the unstable NS into a BH. Although the plateau feature alone can be explained by fallback accretion \citep{fallback_accretion} or a structured jet \citep{structured_jet,structured_jet_lamb}, the following rapid decay in cases such as gamma-ray burst (GRB) 070110 \citep{troja_2007} is too steep to be explained by external shocks, suggesting an internal origin of the emission \citep{Zhang_2006, Zhang_2013, rowlinson_2013}. Additionally, magnetars explain late-time central engine activity well, in particular, flares \citep{Bernardini2015}. In this scenario, observational constraints suggest that the fraction of NS central engines lies at about 15\% to 26\% of Swift sGRB observations \citep{Guglielmi_2024}.

    \cite{duncan&thompson} and \cite{ms_magnetar_first} first proposed that GRBs might be powered by the spin-down of a newly formed and rapidly spinning NS. This statement was initially speculative because the magnetic field of observed pulsars is generally lower than a field that is able to reproduce GRB emission ($B < $ 10$^{14}$~G). However, subsequent observations confirmed the existence of young and highly magnetized NSs \citep{magnetar_evidence_1, magnetar_evidence_2, axp_sd_variation} and tended to consolidate these types of NS as candidates for GRB central engines \citep{magnetar_percentage, Metzger_review, magnetar_central_engine_swift}.

    The physical origin of such a high magnetic field is an active field of research, but it is thought to be at least partly explained by dynamo action \citep{dynamo, Raynaud_2020, Guilet_2022, RS_dynamoMRI_I, RS_dynamoMRI_II, taylor_spruit}. For BNS mergers, the rapid rotation expected to arise in the remnant as a result of the orbital momentum is expected to efficiently amplify the magnetic field \citep{Combi_magnetar_merger, magnetar_merger, B-field_1}. Advances in high-energy astronomy over the past decades have greatly improved our understanding of magnetars, but they remain only partially understood as the equation of state (EoS) of ultra-dense matter or the magnetic field at birth are uncertain. The main implication of the uncertainty on the EoS is that the maximum mass of the nonrotating NS, that is, the Tolman–Oppenheimer–Volkoff limit $M_{\rm TOV}$, is unconstrained.

    Gravitational waves (GWs) offer a complementary and unprecedented perspective on the merger remnant by providing information on progenitor masses and localizations. Interestingly, the GW detection prospects are not limited to face-on sources, while the prompt GRB emission is generally expected to be observable only when the jet is oriented at least partially along our line of sight. If millisecond magnetars are detectable off-axis, the coincident detection in GW and X-rays is arguably the most promising millisecond magnetar detection channel \citep{Zhang_2006, rowlinson_2013}. The first BNS merger detected through GWs, GW~170817 \citep{gw170817_gcn, 170817}, was observed along a weak short GRB, GRB~170817A \citep{grb170817_gcn, grb170817}, and was accompanied by a bright multiwavelength electromagnetic (EM) counterpart, AT2017gfo, identified in galaxy NGC 4993 \citep{170817b, ngc4993}. The transient exhibited rapidly evolving optical and infrared emission consistent with a kilonova powered by the radioactive decay of r-process nuclei synthesized in neutron-rich ejecta, providing strong evidence that BNS mergers are major sites of r-process nucleosynthesis and establishing kilonovae as key EM counterparts to GW events. 
 
    Following the BNS merger, models predict that the EM emission is largely dictated by the properties of the ejected material \citep{kn_review, Metzger_review, Perego_2017, tanaka2017}. As the two NSs inspiral, they become tidally disrupted, causing neutron-rich material to be ejected from the system and dynamically launched into a tail. This ejecta, called dynamical ejecta, has a relatively high mass (10$^{-4} -$ 10$^{-2}$~$M_\odot$; see, e.g., \citealt{dynamical_ej_mass, Nedora_2022, fit_polynomiaux}) and a low electronic fraction, such that $Y_e < $ 0.25. For this reason, the dynamical ejecta is the main site for the formation of r-process nuclei \citep{r_process}. This implies that the dynamical ejecta is lanthanide-rich and heated by radioactive decay. Another implication of the presence of lanthanides is that the dynamical ejecta has a high opacity ($5 \leq \kappa \leq 30 ~\rm cm^{-2}~g^{-1}$) and causes the ``red'' kilonova emission \citep{GW170817_0.1Msol, opacity_calculations}. The mass of the dynamical ejecta greatly depends on the EoS, for example, if the EoS is stiff, the greater pressure entails a larger radius, and thus, a greater number of tidal effects and a higher ejected mass \citep{kn_review, Nedora_2022}.
    The optical transient AT2017gfo requires at least another ejecta component to account for observations, in particular, for the ``blue'' (lanthanide-poor) kilonova emission. This second ejecta component, called post-merger ejecta, is expected to be wind-driven and extracted from the disk after the merger \citep{tanaka2017, Fujibayashi_2020, mass_ejection}. The post-merger ejecta is estimated to have a higher electronic fraction ($Y_e > $ 0.25) than the dynamical ejecta due to the effect of neutrinos. Consequently, the r-process is less efficient in the post-merger ejecta than in the dynamical ejecta, explaining the ``blue'' emission. However, detailed properties of the post-merger ejecta, such as the ejecta mass, electron fraction, and spatial distribution, are still unclear due to the complexity of including neutrino heating, turbulence, and magnetic field in numerical simulations \citep{tanaka2017, Nedora_2022}. This ejecta component is expected to be particularly more luminous in the presence of an NS engine instead of a BH, boosting the luminosity by up to a factor of five \citep{Metzger_review}.
    In addition to the dynamical and post-merger ejecta, based on AT2017gfo observations, other distinct outflow components probably exist that differ with respect to their total mass, velocities and opacities \citep{Perego_2017, mass_ejection}.

    We simulated the GW emission from a population of BNS mergers and the subsequent X-ray light-curve from a millisecond magnetar remnant to infer on our ability to constrain the nature of the merger remnant. One main goal was then to evaluate the detectability of BNS merger events with GW detectors and that of millisecond magnetars with present-day X-ray instruments.

    This paper is structured as follows. Sect. \ref{sect:simu_methods} presents the simulation methods: first to produce the BNS population and their merger GW signal using current and future GW interferometers, and then to estimate the newborn millisecond magnetar X-ray light-curve. Sect. \ref{sect:predictions} gives our simulation results, in particular, our predictions of expected flux, the effect of the magnetar parameters on the detectability, and our suggested observation strategies depending on the observation viewing angle. Finally, in Sect. \ref{sect:simu_discussion}, we discuss the results and review the limitations of our analysis. 
    

\section{Simulation methods}
\label{sect:simu_methods}
Our approach to the simulation of the millisecond magnetar starts with the BNS, of which we build a population, to then simulate their GW emission for a given GW detector configuration; this first step is presented in Sect. \ref{sect:BNS_pop_GW}. Then, for well-localized GW events ($\Delta \Omega$ $\leq$ 50~deg$^2$), follow-up observations to identify an X-ray counterpart are feasible; thus, for these systems, we simulate the light-curve with a millisecond magnetar merger remnant, with methods described in Sect. \ref{sect:grb.py}. Since the emission from the ejecta contains imprints of the nature of the remnant, and dictates the dependence of the emission with the line of sight along with the temporal evolution of the emission, we implement a novel multicomponent ejecta modeling framework. In particular, our approach is novel in that it describes the ejecta as the superposition of multiple components whose properties are determined by polynomial fits to state-of-the-art numerical relativity simulation results \citep{Nedora_2022, angles}, rather than relying on simplified or single-component parameterizations used in previous works. This method also has the merit of creating a self-consistent simulation, as it allows us to adapt the ejecta properties to the BNS orbital parameter and, in particular, to the EoS of the NS. 

\subsection{Synthetic BNS population and GW detectability}
\label{sect:BNS_pop_GW}

First, we set the NS characteristics, namely by choosing a distribution of masses, as described in Sect. \ref{sect:ns_mass_distrib}. The conversion from the relative number of NSs to the expected number of detections is operated thanks to the astrophysical BNS merger rate fixed in Sect. \ref{sect:rates}. Our GW simulation is presented in Sect. \ref{sect:pycbc_intro} for current GW interferometers, and the application to the BNS population is discussed in Sect. \ref{sect:application_O4/5}. 
Sect. \ref{sect:application_et} discusses the case of the next generation of GW interferometers.

\subsubsection{Neutron star mass distribution}
\label{sect:ns_mass_distrib}

Since we aim to accurately reproduce the BNS as detected in GW, we chose to adopt the NS mass distribution based on LIGO Virgo KAGRA (LVK) O3 GW run, as described by \cite{mass_distrib}\footnote{See \cite{Loffredo_2024} for an alternative.}. This distribution is uncertain due to the small sample of observed BNS in GW, as discussed in Sect. \ref{sect:mass_distrib_discussion}.
These authors employ Bayesian inference to find the most probable intrinsic GW NS population properties. 
They show results for two prior models (see Fig. 7 of \citealt{mass_distrib}), called \textit{Power} and \textit{Peak}, as they assume prior mass distribution as a power law and a Gaussian respectively. It is assumed that the masses of the components of the BNS can be independently drawn from the common NS mass distribution. 
For this project, though more precise, considering the two inferred mass distributions would have been too demanding in terms of computational resources and time. \cite{mass_distrib} find that the \textit{Power} and \textit{Peak} population models are consistent within the 90\% credible interval. Additionally, integrating under both curves yields a less than 5\% difference on the distribution of predicted mass. We thus chose the \textit{Peak} distribution to simulate each NS component mass of the GW BNSs, which slightly favors the lower mass NSs in which we are interested.

\subsubsection{Binary neutron star merger rate}
\label{sect:rates}
To estimate the rates of detection from our number of detected systems, we computed
\begin{equation}
\label{eq:rate}
    n = \frac{N_{det}}{N_{tot}} \times R \times \mathcal{V},
\end{equation}
with $n$ the predicted rate of detectable magnetar (yr$^{-1}$), $N_{det}$ our number of GW and/or EM detections within a given sample, $N_{tot}$ the total number of BNS simulated, $R$ the BNS merger rate (Gpc$^{-3}$ yr$^{-1}$), and $\mathcal{V}$ the simulated volume (Gpc$^{3}$). In Eq. \eqref{eq:rate}, the astrophysical BNS merger rate~$R$ is a very uncertain term, difficult to estimate with a modeling approach, as the formation of the binary depends on a number of unconstrained parameters, namely the efficiency of the common envelope phase \citep{common_enveloppe_efficiency}. Observationally, the evaluation of the merger rate is also uncertain, as the population of observed BNS that have merged consists of only two systems. 
Using Bayesian inference on the LVK O3 GW catalog with a number of prior models, \cite{mass_distrib} conclude that the BNS merger rate is comprised in the range 10 $-$ 1700 Gpc$^{-3}$ yr$^{-1}$. 
From the fact that no BNS merger has been confidently claimed from O4 data so far (despite the increased BNS range compared to the previous observing run), \cite{factor_bns_range} infer a merger rate of 7.6 $-$ 250~Gpc$^{-3}$.
Using a modeling approach, \cite{et_data} find a merger rate of 23 $-$ 107~Gpc$^{-3}$ yr$^{-1}$ for a common envelope efficiency $\alpha$ of 0.5 and 1.0, respectively. 
Since this modeled rate is consistent with the LVK O3 results, we provide the number of detection per year assuming a $R = $~100 Gpc$^{-3}$~yr$^{-1}$ merger rate. 

\subsubsection{GW simulation for LVK O4 and O5}
\label{sect:pycbc_intro}
We compute a GW merger signal for BNS sources using the \texttt{PyCBC} software. \texttt{PyCBC} \citep{pycbc} is a free and open software to study GWs. The main aim of this simulation code is to generate skymaps, containing information on the localization precision of the source, based on given BNS parameters, and according to a given interferometer configuration. The process of the GW detection simulation can be summarized in four steps:
the setting of the BNS parameters, the waveform simulation, the noise simulation, and the skymaps generation. 
The initial step requires specifying the intrinsic parameters of the system. The masses of each binary component and their distance are sampled from a BNS population as described in the next subsection. 
Spins were picked uniformly in the range $[-0.05 ;  0.05]$ and the eccentricity was set to zero for simplicity and to optimize computation time. Though tidal effects are accounted for in our EM predictions (see Sect. \ref{sect:grb.py}), for the purpose of GW detectability and sky localization, we chose to neglect them in our GW simulation.
Sky location and system inclination are sampled from isotropic distributions, i.e., uniform in right ascension, in $\cos \delta$, and in $\cos i$, with $\delta$ the declination and $i$ the binary system inclination.
 
\begin{figure} 
    \centering
    \includegraphics[scale = 0.5]{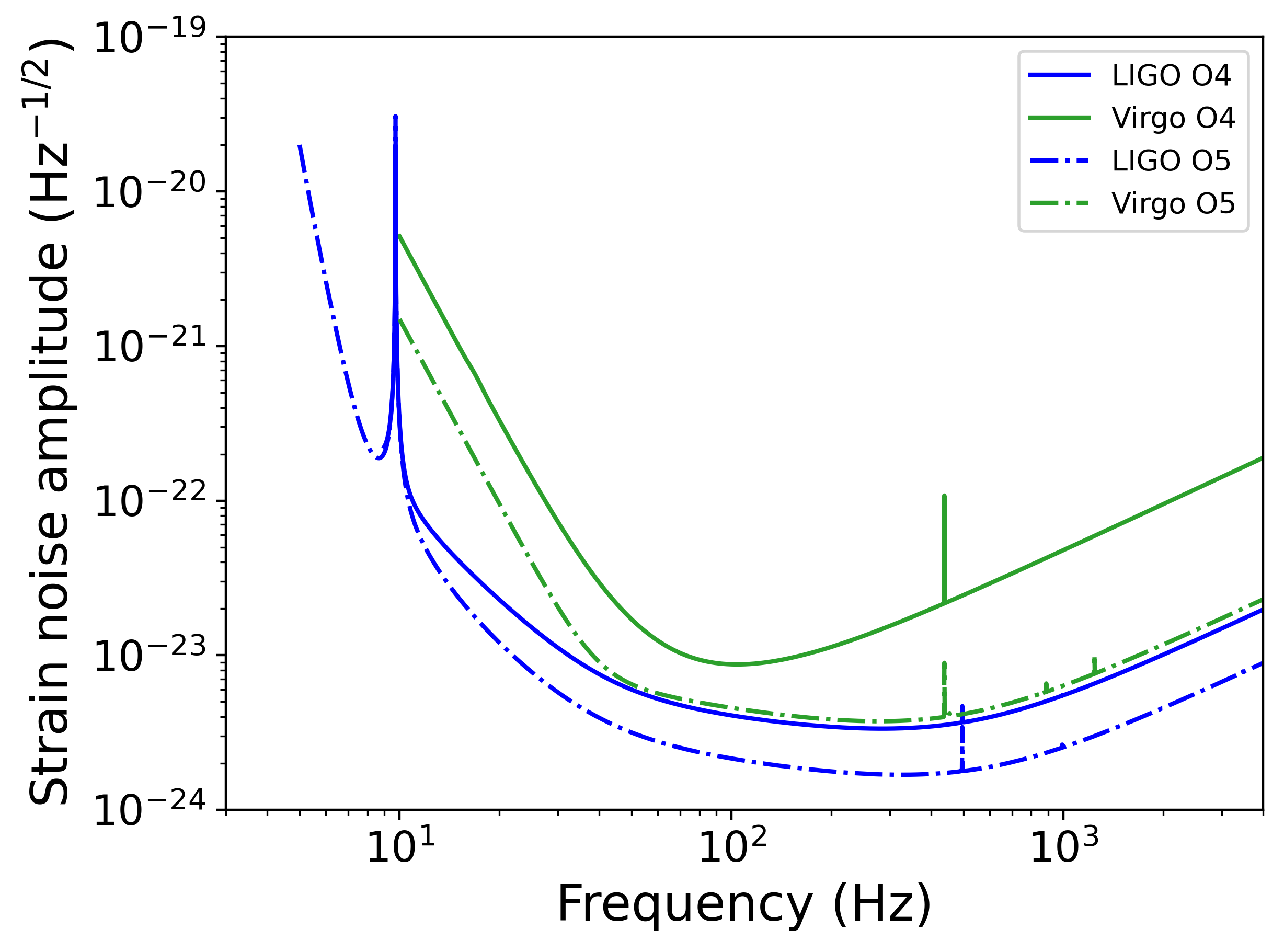}
    \caption{Amplitude spectral density curves (i.e., strain noise amplitude as a function of frequency) chosen for the GW simulation depending on the configuration. O4 ASD curves are shown as filled lines, and O5 ASD curves are shown as dashed lines for the LIGO (blue) and Virgo (green) instruments. 
    }
    \label{psd}
\end{figure}

To reproduce the detected GW signal from a given BNS, we must first estimate the waveform (i.e., the strain as a function of time) produced by the BNS inspiral and merger. The ring-down signal is not detectable by current GW interferometers (though it might be detected by the next generation of GW interferometers; see Sect. \ref{sect:obs_conditions_discussion}), and as such has not been modeled here. We chose the \textit{TaylorF2} waveform model, as it is a simple model that captures the essential physics of a BNS merger for the purpose of determining its detectability and sky localization with LVK detectors.
The noise in a GW signal is the main driver of the interferometer detection capability. The statistical properties of this noise are modeled using the amplitude spectral density (ASD).
The ASD is dependent on the detector configuration, in particular on the instruments technical upgrades. It was therefore adapted to reproduce different actual detector sensitivities we simulated. We used common ASD Models for GW detectors provided by \texttt{PyCBC}. More precisely, to reconstruct O4 sensitivity, we configured a 175~Mpc range \citep{factor_bns_range} for the two LIGOs (with the ASD name aLIGOaLIGO175MpcT1800545), and we chose an advanced O3 configuration for Virgo with low sensitivity (with the ASD name AdVO3LowT1800545) to reconstruct the instrument 50$-$55~Mpc range. Since KAGRA operated only for a short period of time and over a short BNS range, we did not account for this interferometer in our O4 simulations. For O5, we chose the Advanced LIGO design sensitivity (named aLIGOZeroDetHighPower) for the two LIGO instruments, corresponding to a BNS range of 325~Mpc, and the Virgo design sensitivity (named AdVDesignSensitivityP1200087), corresponding to an approximate BNS range of $\sim$ 145~Mpc, with the addition of KAGRA (KAGRA128Mpc) with an optimistic 128~Mpc range. The O5 starting dates, duration, and sensitivities are based on the current best assumptions, and will likely be adjusted as the start of the observing run approaches. The ASD curves for LIGO and Virgo in the O4 and O5 configurations are shown in Fig.~\ref{psd}.

From the detector signal, i.e., the BNS GW signal with the addition of the interferometer noise, the signal-to-noise ratio (S/N) is computed. If the S/N in one interferometer exceeds 5.5, a skymap can be produced. All interferometers above this S/N threshold are included in the localization input. This step is not operated directly by \texttt{PyCBC}, and calls for the public software package \texttt{BAYESTAR} \citep{bayestar}.  
The 50\% and 90\% credible region are computed, i.e., the smallest area containing 50\%, respectively 90\%, of the posterior probability.

\subsubsection{Application to the BNS population for current GW interferometers}
\label{sect:application_O4/5}

To build the synthetic BNS population, each NS component mass is randomly chosen from the distribution described in Sect. \ref{sect:ns_mass_distrib}, and the distances are picked up to a limit distance dependent on the considered GW detectors' range. This defines a number of BNS systems, uniformly distributed within a sphere with the Earth at its center. 

\begin{figure}
    \centering
    \begin{subfigure}{.45\textwidth}
        \centering
        \includegraphics[width=1.\linewidth]{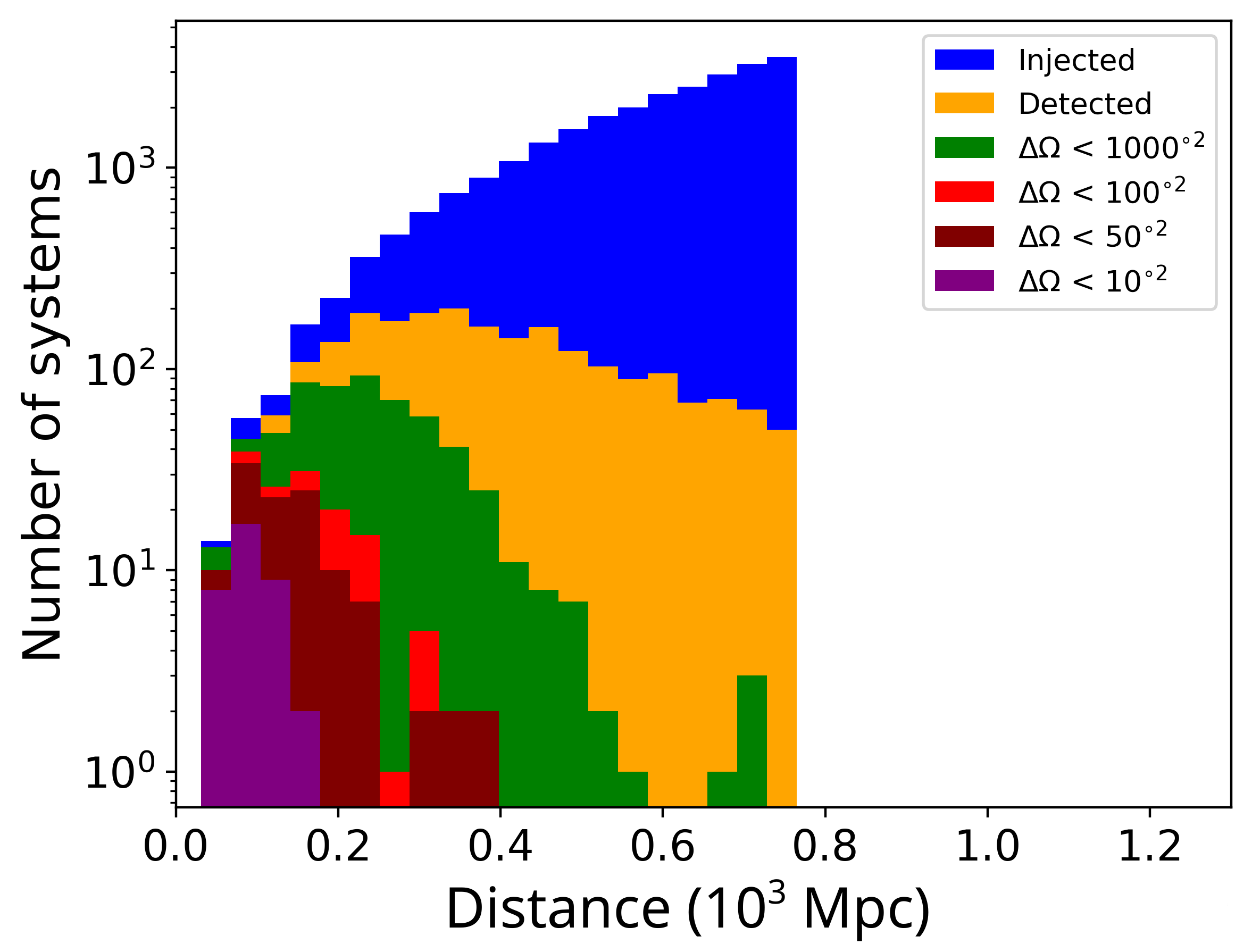}
    \caption{}
    \end{subfigure} 
    \begin{subfigure}{0.45\textwidth}
        \centering
        \includegraphics[width=1.\linewidth]{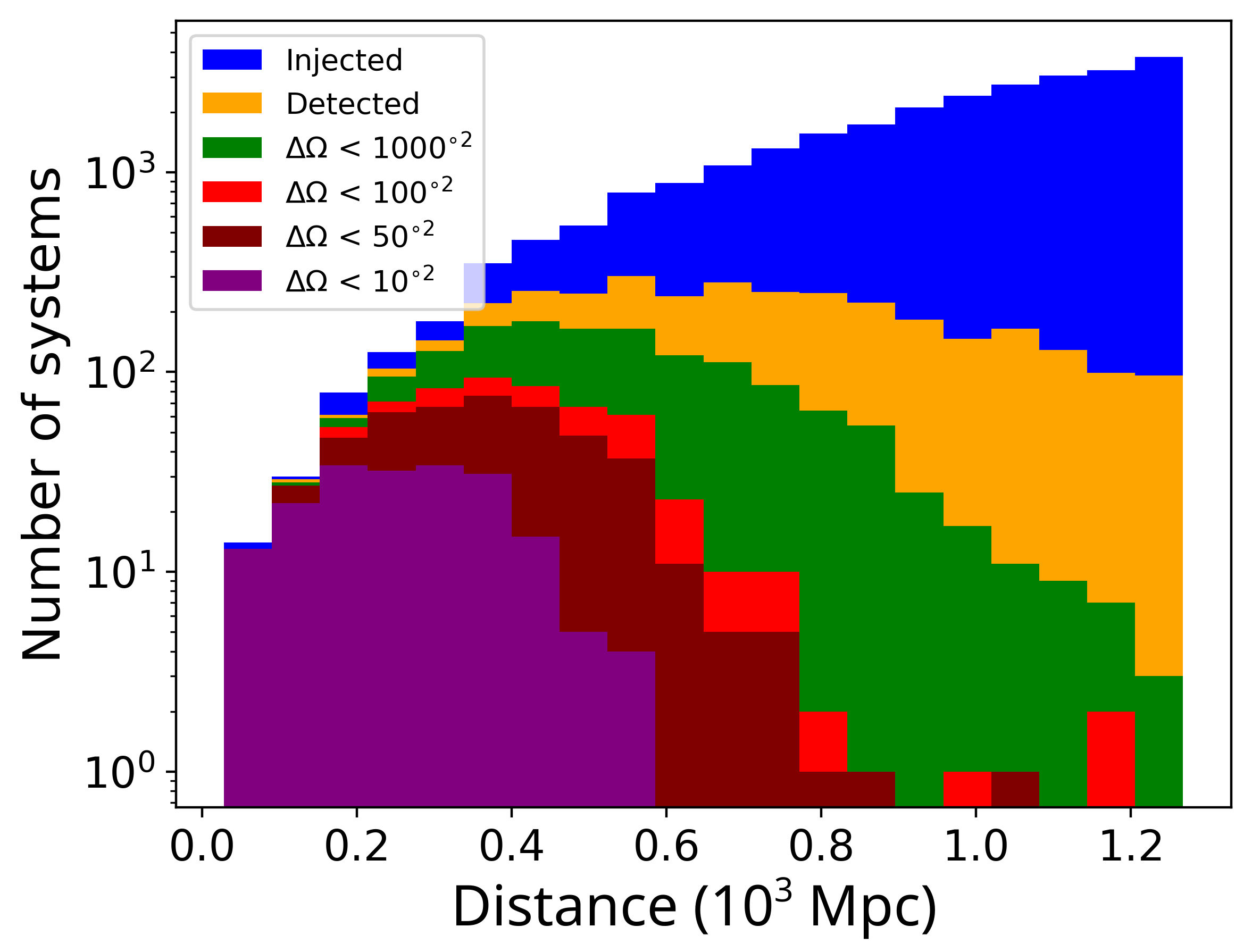}
    \caption{}
    \end{subfigure}
    \caption{Distribution of the GW localization accuracy obtained on the O4 (panel a) and O5 (panel b) GW run simulations. All the BNS of the synthetic population that went through the GW detection pipeline are shown in blue, all detected systems are shown in orange, and the localized systems are shown with colors dependent on their localization accuracy (noted here $\Delta \Omega$). }
    \label{errorboxes_O4_O5}
\end{figure}

We ran simulations on two sizes of sphere for each GW configuration: first on a sphere wide enough to comprise all BNS detectable by the GW interferometers no matter their inclination, then on a narrower sphere to get sufficient statistics on well-localized systems.  
For the larger sphere simulations, we chose a sphere of radius $D_{max} =$~765~Mpc for O4 and $D_{max} =$~1275~Mpc for O5, estimated based on LIGO BNS range multiplied by a factor three to account for all inclinations and sky locations of the source\footnote{A factor 2.3 is the amplitude of variations expected over all possible inclinations, but we chose a larger factor to account for favorable interferometer antenna patterns, i.e., when a GW is incoming in such a way that a maximum interferometers arms deformations is induced and the signal is seen with maximum S/N.} \citep{factor_bns_range}. These simulations yielded the distributions of GW localization accuracy results as a function of distance, as shown in Fig. \ref{errorboxes_O4_O5}. From this, we ran simulations on smaller spheres of radius $D_{zoom} = $~300~Mpc and  $D_{zoom} = $~800 Mpc for O4 and O5 respectively, as they encompass at least 95\% of systems with a localization accuracy of 50~deg$^2$ or smaller.

\subsubsection{Application to the BNS population for the next generation of GW interferometers}
\label{sect:application_et}
Concerning the next generation of GW detectors, as of today, two projects stand out. 
\textit{Einstein} Telescope \citep[ET;][]{et}
is an ambitious proposal to construct an underground third-generation GW interferometer.
It will operate on a wider frequency band (10$^{0}$~Hz $-$ 10$^{4}$~Hz, while the LVK networks operates best between 10$^{1}$~Hz $-$ 10$^{3}$~Hz). This European project is foreseen to begin observations in $\sim$ 2035, with an operation duration of $\sim$ 50 years. The configuration, and thus, the final performance, of this instrument is still not settled. The standard configuration is a triangle of 10~km arms, but two L-shaped interferometers of 15~km arms is another configuration under study.
Cosmic Explorer (CE; \citealt{ce}) is a comparable project to ET, but mostly US-funded. The design concept for Cosmic Explorer features two facilities, each housing a single L-shaped interferometer.
Each CE facility will have two 40~km (20~km for the second site) ultrahigh-vacuum beam tubes, roughly 1~m in diameter, built in an L-shape on the surface of flat and seismically quiet land in the US. 
We considered the contribution of ET in a triangle configuration, along with only the first CE site, for an optimistic but realistic performance of the future generation of GW detectors. 

\begin{figure}
    \centering
    \begin{subfigure}{.45\textwidth}
        \centering
        \includegraphics[width=1.\linewidth]{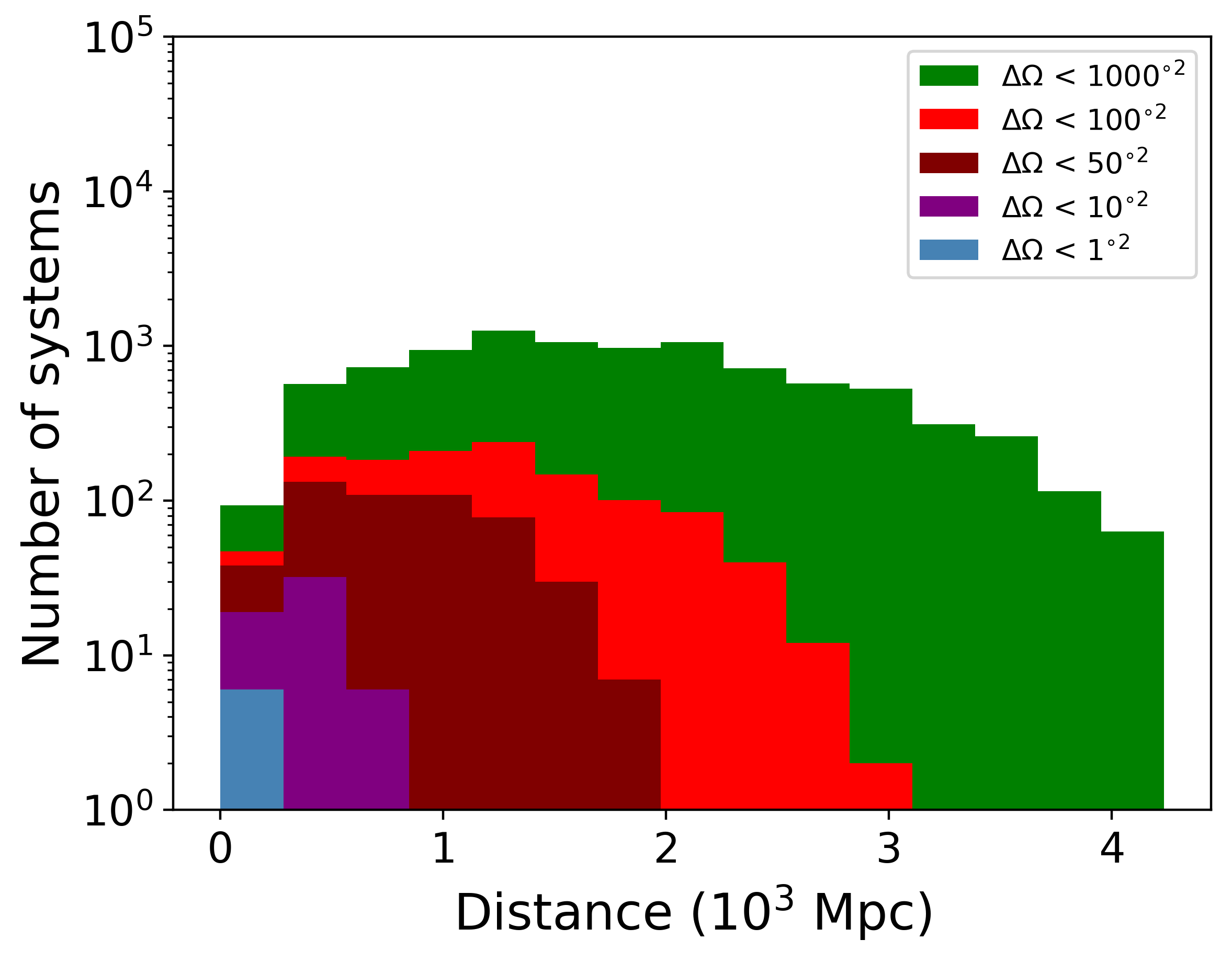}
    \caption{ET $\Delta$ 10~km.}
    \end{subfigure} 
    \begin{subfigure}{0.45\textwidth}
        \centering
        \includegraphics[width=1.\linewidth]{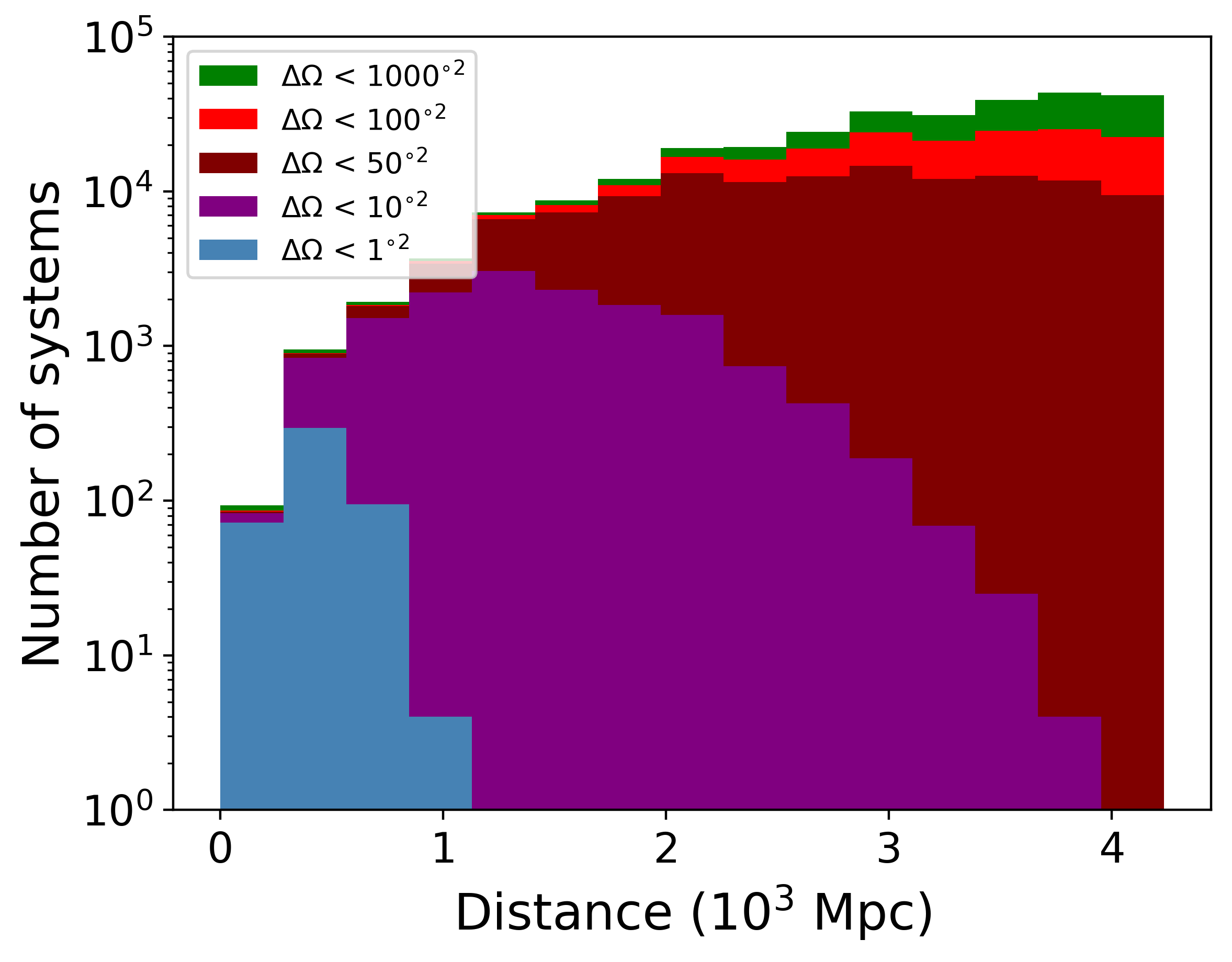}
    \caption{ET $\Delta$ 10~km + 1 CE 40~km.}
    \end{subfigure}
    \caption{Distribution of the GW localization accuracy from \cite{et_data} data on two ET configurations: a triangle of 10~km arms by itself (a), or with the one CE interferometer (b). The localized systems are shown with colors dependent on their localization accuracy (noted $\Delta \Omega$).}
    \label{erroboxes_triangles}
\end{figure}

The simulation of ET and CE performance is complex and outside the scope of this paper. Instead, we made use of literature results. \cite{et_data} provide public catalogs containing all the parameter estimations for 10 years of observations of BNSs up to a redshift equal to 1, using the ET alone or in a network of current or next-generation detectors. The localization accuracy distribution for each of these configurations is shown in Fig. \ref{erroboxes_triangles}. Concerning the NS mass distribution, they draw the component masses of the NS binaries from two different mass distributions: Gaussian or uniform. To reproduce the input distribution of mass as closely as for LVK simulations (see Sect. \ref{sect:ns_mass_distrib}), we chose ET GW simulations with the uniform mass distribution. The uniform mass distributions range in 1.1~M$_\odot\ - $ $M_{max}$, where $M_{max}$ depends on the selected EoS: either the BLh EoS \citep{eos_blh} with $M_{max} = $ 2.1~M$_\odot$, or the APR4 EoS \citep{eos_apr4} with $M_{max} = $ 2.2~M$_\odot$. Once again to stay consistent with previous GW simulation, we chose the BLh EoS, yielding a uniform mass distribution ranging from 1.1~M$_\odot\ - $ 2.1~M$_\odot$ for the input BNS synthetic population. Compared to the O4 and O5 chosen NS mass distribution, spanning $\sim$ 1.12~M$_\odot\ -$ 1.96 ~M$_\odot$, this distribution of mass extends to somewhat larger masses. 
This means that results for GW simulation methods are not strictly equivalent: for the same considered merger rates and GW detector configuration, the mass distribution considered for ET induces less millisecond magnetars than the mass distribution considered for O4 and O5. This is confirmed by performing a simplified analysis on ET data: though the fraction of stable NS is consistent within error bars, the fraction of millisecond magnetar candidate (see Sect. \ref{sect:critical_period} for a precise definition) decreases from 35\% to 30\% for a high $M_{\rm TOV}$ EoS considering this distribution of mass only.

Last, we note that longer GW detected signals also imply that some BNS might be detected before the merger. For these BNS, earlier EM observations can be considered, impacting the observation strategy and detection prospects. Cosmological redshift also impacts the emitted signals duration and wavelength. This was accounted for in the simulated data presented here; our methods to reproduce the effect of cosmological redshift on the remnant X-ray emission are presented in Appendix \ref{ap:z_effect}.

\subsection{X-ray light-curve simulation}
\label{sect:grb.py}

At present, X-rays are the only emission capable of constraining the nature of the remnant from a BNS merger. As such, realistic simulation of the X-ray emission of the remnant is essential for detection predictions. We use the simulation code \texttt{GRB} originally developed at CEA-Saclay by J. Guilet, R. Raynaud, and M. Bugli.
The aim of this Python code is to produce light-curves associated with the coalescence of two NSs, assuming that the compact object formed is a magnetar that loses energy through magnetic braking induced by the dipole component of the NS magnetic field. It uses a model adapted from the work of \cite{Sun}, but also from \cite{Gompertz} and \cite{energy_injection}. For further information on the simulation code, documentation can be found on the \texttt{GRB}  \href{https://github.com/rraynaud/GRBs.git}{GitHub repository} (Commit number 
2350f99).

\subsubsection{Observation geometry}
\label{sect:obs_geom}
We assume that the millisecond magnetar produces an isotropic X-ray emission \citep{Zhang_2013}. 
The reprocessing of this emission by the ejecta makes expected observations dependent on the observation angle and time \citep{tanaka2017, Perego_2017, kn_review}. The ejecta will first absorb this radiation in certain directions and then re-emit it as a blackbody spectrum. 
Computing the light-curves therefore requires determining the dynamic evolution of the ejecta. For this, we use the model of \cite{ejecta_dynamics}, which reproduces the energy injection by the central magnetar along with an additional heating by radioactive element decay. 

\begin{figure}
    \centering
    \includegraphics[trim={340 60 0 0},clip,width=1.\linewidth]{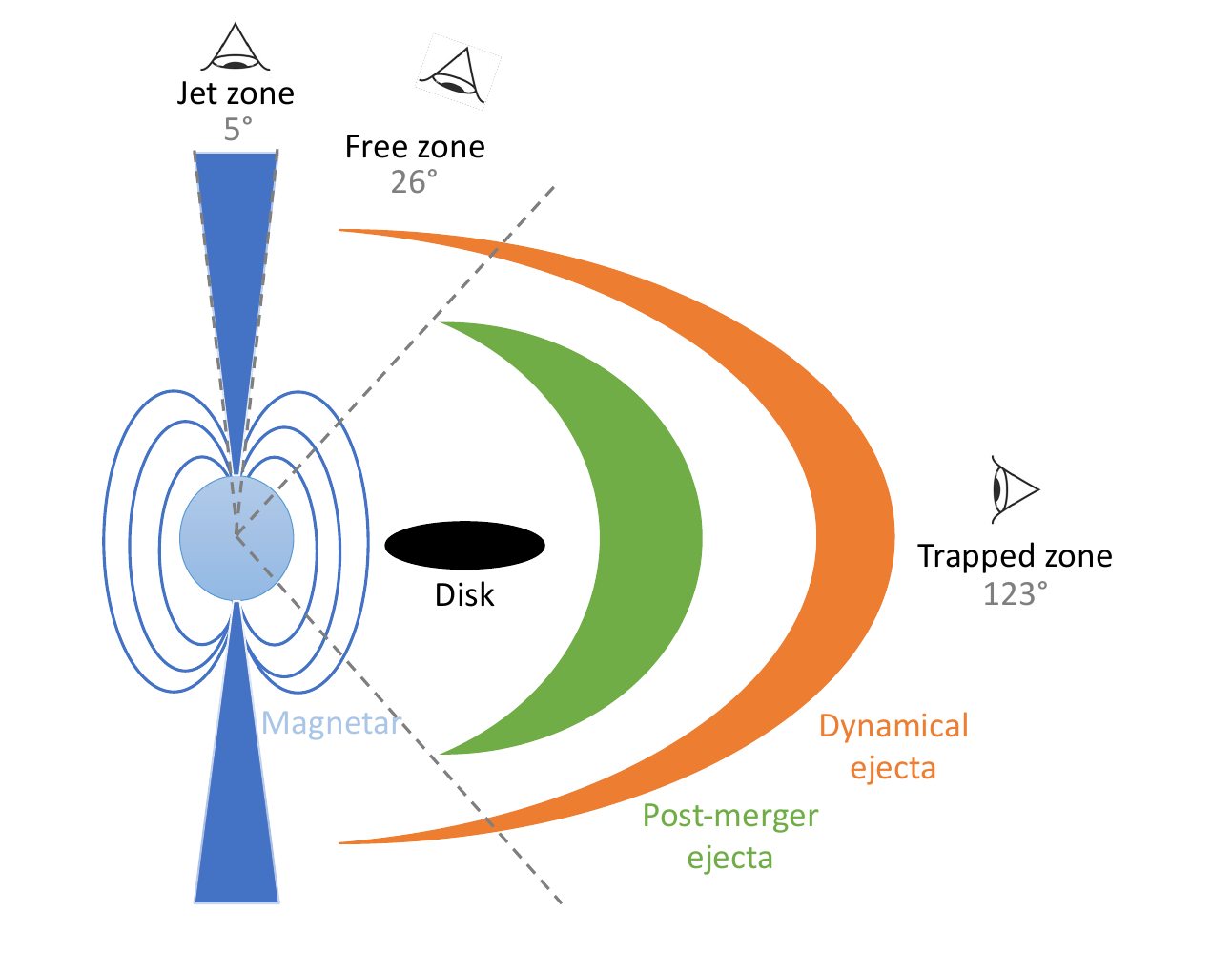} 
    \caption{Schematized ejecta geometry seconds after the merger. A fraction of the disk goes into post-merger ejecta. The dynamical ejecta is extracted at the time of the merger and expands rapidly. The different emission zones are delimited by dashed lines, and we report the corresponding opening angles. }
    \label{obs_geom}
\end{figure}

Our ejecta model is schematized in Fig. \ref{obs_geom}. The jet zone, although with the smallest solid angle (the jet aperture is estimated to be around $5^\circ$ \citep{radio_structured_jet}) is the configuration in which most mergers have been observed so far, in the form of sGRBs.
Outside the jet, we follow \cite{Sun} in dividing the ejecta into two distinct zones: the trapped zone around the equator and the free zone closer to the jet axis. In \cite{Sun}, only the dynamical ejecta was modeled, assuming it was absent in the free zone. However, it is probable for the dynamical ejecta to be present in the free zone, but with a density lower than in the trapped zone. To model this, we added $1/10$ of the dynamical ejecta mass on the free zone, based on \cite{free_opening_angle} three-dimensional dynamical-spacetime general-relativistic magnetohydrodynamic simulations results with a $B_{dip} = 10^{15}$~G hypermassive magnetar as a remnant. From \cite{free_opening_angle}, we also estimate the uncertain opening angle of the free zone at 26$^\circ$. This leaves a 123$^\circ$ wide opening angle for the trapped zone, which therefore covers the largest fraction of solid angle. 

To have a more realistic evolution of the luminosities, we used polynomial fits informed by numerical simulations to estimate the masses and opacities of the ejectas, main drivers of the millisecond magnetar received emission. 
This permits a novel, more complex and physically accurate ejecta modeling through three main points: (i) a more precise and system-dependent estimation of the ejecta masses, (ii) the inclusion of the post-merger ejecta, and (iii) the determination of the opacity of the ejecta based on electronic fraction estimations. In addition to allowing us to access more physically accurate values of the opacity, this method also has the merit of creating self-consistent simulations, as it allows us to adapt the ejecta properties to the BNS orbital parameter and to the EoS.

Firstly, considering its forming conditions, the dynamical ejecta properties, and in particular, its mass, are directly connected to tidal effects. Consequently, the dynamical ejecta mass can be evaluated thanks to fitting formulae based on numerical simulations results. 
\cite{Nedora_2022} derive polynomial fits from a large sample of numerical relativity simulations (including neutrino transport) over various EoS.
These polynomial fits depend on the mass ratio of the BNS $q$ and the reduced tidal parameter $\Lambda_t$. The mass ratio of a BNS is defined as the ratio of each BNS component masses $q = M_1/M_2$ with $M_{i=1,2}$ the NS masses and $M_1>M_2$.
The reduced tidal deformability $\Lambda_t$ quantifies the tidal effect during the inspiral and is expressed as follows:
\begin{equation}
\label{eq:lambda_t}
\Lambda_t = \frac{16}{13} \left(\frac{(M_1 + 12 M_2) M_1^4 \lambda_1}{(M_1 + M_2)^5} + 
\frac{(M_2 + 12 M_1) M_2^4 \lambda_2}{(M_1 + M_2)^5} 
\right),
\end{equation}
with $\lambda_i$ the quadrupolar tidal parameters of the NS component $i=1,2$ of the BNS. To evaluate the quadrupolar tidal parameters for each NS, we use CompOse, an online service that provides data tables for different EoS \citep[see][]{compose}. Firstly, for a grid of NS masses, we solve the TOV equations to calculate the tidal deformability for each EoS \citep{HindererTidaldeformability}, and then we interpolate to obtain the mass-tidal deformability relation for each NS.

The post-merger ejecta, extracted from the remnant disk, is also expected to be present on the line of sight in the trapped zone configuration (see Fig. \ref{obs_geom}).
We estimate that a fraction $f_{ej}$~=~0.4 of the disk is extracted to form the post-merger ejecta (see \cite{mass_ejection, magnetar_merger}, or alternatively \cite{post_merger_ejecta} evaluated $f_{ej}$ at 0.3). To evaluate the disk mass, we also used polynomial fits from \cite{Nedora_2022}. 
 
Then, to get the ejectas opacity ~$\kappa$, we linked it to the ejectas electronic fraction~$Y_e$, that we were also able to estimate thanks to polynomial fits. 
Since it is due to absorption by atoms and ions, the dominating bound-bound opacity is complex to compute: one has to consider all possible transitions from each present chemical element. 
\cite{opacity_calculations} considered oscillator strength of most probable spectral lines by performing systematic atomic structure calculations of r-process elements. 
They found the average opacities for the mixture of r-process elements to be $\kappa \sim$ 20 $-$ 30~cm$^2$ g$^{-1}$ for an ejecta component at the electron fraction of $Y_e$ < 0.20, $\kappa \sim$ 3 $-$ 5~cm$^2$ g$^{-1}$ for $Y_e$ = 0.25 $-$ 0.35, and $\kappa \sim$ 1~cm$^2$ g$^{-1}$ for $Y_e$ = 0.40. Since no values were specified for $Y_e$ = 0.20 $-$ 0.25, we performed a linear interpolation to obtain the opacity over those values. 

Radiation at the moment of the merger is initially blocked by the optically thick dynamical ejecta in the free zone, and by the post merger ejecta as well in the trapped zone. However, photons may eventually escape as the ejecta expands and cools, thus becoming optically thin. 
Thus, the magnetar radiation produced through the magnetic dipole luminosity $L_{\text{dip}}$ has been partially absorbed by the ejecta of a certain optical depth $\tau$, and is scaled by the efficiency parameter~$\eta_{\text{dip}}$ of conversion of the magnetar rotational energy into X-rays, making the wind X-ray luminosity component,
\begin{equation}
   L_{\text{X,wind}} = \eta_{\text{dip}} e^{-\tau} L_{\text{dip}},
\end{equation}
where we chose $\eta_{\text{dip}} = 0.01$ (see Discussion \ref{sect:param_degen_discussion}). The optical depth $\tau$ is computed from the opacity $\kappa$ and the evolution of the ejecta $\tau = \kappa (M_\text{ejecta}/V')(R/\Gamma)$, where $V'$ is the comoving volume \citep{Sun}.

Finally, the dynamical ejecta itself also emits due to the cooling of radioactively heated material, adding the blackbody ("bb" hereafter) component to the X-ray luminosity \citep{Sun}. The blackbody component results from the integration over a frequency range of Planck's emission law,
\begin{equation}
   I_{\text{bb}} = \frac{8(\pi \Gamma R)^2}{h^3 c^2 \nu} \frac{({h \nu}/{\Gamma})^4}{e^{\frac{h \nu}{\Gamma k_\text{B} T}}-1},
\end{equation}
where $\Gamma$, $R$, $T$, $k_\text{B}$, $h$, $c$, and $\nu$ are the Lorentz factor, the NS radius and the temperature of the ejecta, the Boltzmann and Planck constants, the speed of light, and the frequency of the electromagnetic emission, respectively. 

\subsubsection{Spin-down}
\label{sect:sd}
We considered an NS with a radius $R$, a mass $M$, a moment of inertia $I$, and a surface dipolar magnetic field $B$, spinning with angular frequency $\Omega$. We also consider the magnetar to be surrounded by an accretion disk, formed by matter falling back onto the magnetar after coalescence (known as “fallback accretion”). Then, following interaction with the powerful magnetic field of the magnetar, this matter can, under certain conditions, be accelerated to super-Keplerian speeds. This results in the ejection of matter, as the centrifugal force overcomes the gravitational attraction of the magnetar. To simulate this, and the impact of the so-called ``propeller'' effect on the luminosity, we adopt the formalism used by \cite{Gompertz}, as described in Appendix \ref{ap:propeller_regime}. Then finally, as mentioned previously, the magnetar draws energy from its rotational energy reservoir,
\begin{gather}
          E(t) = \frac{1}{2} I(t) \Omega(t)^2.
    \label{eq:energy_reservoir}
\end{gather}

The time evolution of the magnetar angular frequency $\Omega$ and moment of inertia $I$ are determined by the sum of the torques, composed of the magnetic dipole, gravitational, and accretion torques ($N_{\text{dip}}$, $N_{\text{grav}}$ and $N_{\text{acc}}$, respectively),
\begin{gather}
      \frac{\partial (I \Omega)}{\partial t} = \sum N = N_{\text{dip}} + N_{\text{grav}} + N_{\text{acc}}.
      \label{eq:N_Omega_dt}
\end{gather}
The gravitational torque comes from the emission of GW in the case of nonzero ellipticity in the remnant. The accretion torque can be positive and negative depending on the accretion and propeller regime of the disk. 

The derivative of Eq.~\eqref{eq:energy_reservoir} yields the magnetar spin-down,
\begin{gather}
      \Dot{E} = I \Omega  \Dot{\Omega} + \Dot{I} \frac{\Omega^2}{2}.
    \label{eq:E_budget_expanditures}
\end{gather}
The second term is due to decrease of the NS moment of inertia as the rotation slows down. Our approach to the implementation of the time evolution of the NS moment of inertia, not featured in other models such as in \cite{Sun}, is described in next section (Sect. \ref{sect:I}). In the energy evolution, the magnetar loses energy due to its emission that can be described by the following luminosities \citep{merger_lum, Usov1992, energy_injection_2}:

\begin{equation}
\label{eq:rate_deceleration}
   L(\Omega) = -L_{\text{sd}} - L_{\text{GW}},
\end{equation}
where $L_{\text{sd}}$ is the spin-down luminosity, resulting from the contributions of the dipolar and propeller luminosities, 

\begin{align}
L_{\text{sd}} &= L_{\text{dip}} + L_{\text{prop}}.
\end{align}

The first term of this equation corresponds to the slowdown caused by the magnetic torques. Then, noting $\mu = B R^3$ the magnetic moment, we estimate the dipole luminosity as
\begin{equation}
    L_{\text{dip}} = N_{\text{dip}} \Omega = \frac{\mu^2 \Omega^4}{6 c^3} = 9.6 \times 10^{48}\ B^2_{15}R^6_{6}P^{-4}_{-3} \ \mathrm{erg\ s^{-1}},
    \label{L_dip}
\end{equation}

with $B_{15} = B/10^{15}~G)~$ and $P_{-3}$ the NS rotation period in milliseconds. The propeller luminosity may also add to the NS X-ray emission, depending on the propeller or accretion regime of the disk (see Appendix \ref{ap:propeller_regime}), but to a lesser extent than the propeller luminosity, and typically ending before $~10^2$ seconds (see Sect. \ref{sect:light_curves}), once the accretion disk is consumed. To model the propeller luminosity, we adopted the same formalism as \cite{Gompertz},
\begin{align}
L_{\text{prop}} &= -\left[ N_{\text{acc}} \Omega + \frac{G M_* \dot{M}}{r_m} \right].
\end{align}

Finally, the emission of GW, corresponding to the second term of Eq. \ref{eq:rate_deceleration}, is expressed as 
\begin{equation}
   L_{\text{GW}} = \frac{32GI^2 \epsilon^2 \Omega^6}{5 c^5} = 1.1 \times 10^{41}\ I^2_{45}\epsilon^2_{-2}P^{-6}_{-3} \ \mathrm{erg\ s^{-1}} ,
\end{equation}
where $\epsilon$ is the ellipticity of the NS.
This luminosity is particularly prevalent at times close to the merger, where instabilities such as the bar-mode instability \citep{Gompertz}, or locally high magnetic field, may arise. Nevertheless, the GW luminosity is of relatively low amplitude, $1.1 \times 10^{41}\mathrm{erg\ s^{-1}}$ with our chosen parameters, corresponding to a $h \sim 2.6 \times 10^{-25}$ GW strain for an NS remnant at 100~Mpc \citep{post_merger_strain}. Such a strain is one order of magnitude below LVK(I) current detection abilities (see Fig. \ref{psd}), consequently we consider this GW signal undetectable at present.

\subsubsection{Evolution of the moment of inertia}
\label{sect:I}
In Eq. \eqref{eq:E_budget_expanditures}, the second term accounts for the evolution of the magnetar moment of inertia, which is expected to change as the magnetar slows down \citep{I_decrease}. Consequently, the moment of inertia directly impacts the magnetar luminosity, and the time of its potential collapse into a BH, which makes it a crucial parameter here. We propose the following equation to define the behavior of $I$ with rotation rate, 
\begin{gather}
      I = I_{\rm norot} + \left(\frac{\Omega}{\Omega_0}\right)^n \left(I_{\rm rot} - I_{\rm norot} \right),
      \label{eq:I_def}
\end{gather}
where $I_{\rm rot}$ is the moment of inertia of a maximally spinning NS (highest mass and highest spin possible, extracted from \citealt{data_EoS}), $I_{\rm norot}$ is the final moment of inertia of the NS, $n$ is a configurable index to describe the dependence of the moment of inertia with the angular velocity, and $\Omega_0$ is the initial angular frequency.
Once the NS slowed down significantly, we approximate the final moment of inertia $ I_{\rm norot}$ as one of a uniform sphere, $I_{\rm norot} = 0.35 M R^2$ \citep{Gompertz}.
Based on the $\Omega^2$ dependence of the centrifugal force, we made the simplifying hypothesis that $n$=2. 
Injecting Eq. \eqref{eq:I_def} with $n=2$ in the time evolution of the angular frequency (Eq. \eqref{eq:N_Omega_dt}) yields 
\begin{gather}
          \frac{\partial \Omega}{\partial t} =  \frac{\sum N}{I + 2 \left(\frac{\Omega}{\Omega_0}\right)^2  \left(I_{rot} - I_{\rm norot} \right)}.
          \label{eq:omega_evol}
\end{gather}

Based on our definition of the time-evolving moment of inertia (Eq. \eqref{eq:I_def}), the evolution of $\Omega$ is described by
\begin{gather}
             \frac{\partial \Omega}{\partial t} =  \frac{\sum N}{I + 2 \left(I - I_{\rm norot} \right)}.
\end{gather}

\subsubsection{Potential collapse into a black hole} 
\label{sect:critical_period}
As mentioned before, the NS remnant is susceptible to collapse into a BH. The instant of collapse, if any, is dictated by the mass of the binary and the EoS. We define the mass of the remnant $M$ by subtracting an estimated ejected mass during the merger, which based on AT 2017gfo observations, we evaluate as $M = M_1 + M_2 - 0.1\ M_\odot$ \citep{GW170817_0.1Msol}. If the remnant has a mass below the maximum mass of the nonrotating NS $M_{\rm TOV}$ for a given EoS, the central engine is a stable NS, which will continue to exist after the merger and associated emissions. If, on the other hand, $M > M_{\rm TOV}$, its evolution is directly linked to the evolution of the rotation profile. 
Indeed, the remnant can be a supermassive NS (SMNS), i.e., stabilized by solid rotation, if the mass of the remnant is such that $M< M_{\rm max}$, where $M_{\rm max}$ is the maximum mass of an NS in solid body rotation. Now, for hypermassive NSs (HMNSs), arising if $M_{\rm max}< M< M_{\rm th}$, where $M_{th}$ is the threshold mass for the prompt collapse of the remnant to a BH, the remnant is stabilized by differential rotation. 

Here, we do not consider HMNSs; this is because they are expected to be shorter lived, i.e., with a $\sim$ 100~ms lifetime due to the rapid loss of differential rotation. SMNS spin-down takes place through less efficient processes (e.g., magnetic dipole radiation or GW emission), implying it is likely to survive on longer timescale than HMNS \citep{kn_review}, making them potential targets for X-ray follow-up observations. We therefore treated any remnant with a mass such that $M > M_{\rm max}$ as a BH and did not consider it for the \texttt{GRB} code.
 
The maximum mass a SMNS can support can be parametrized as a function of its rotation period $P$ and the EoS $M_{\rm TOV}$ \citep{eq_NS, ns_collapse_2},
\begin{equation}
    M_{max} = M_{\rm TOV}(1 + \alpha P^{\beta}),
    \label{masse_max}
\end{equation}
with $\alpha$ and $\beta$ phenomenological parameters that, along with $M_{\rm TOV}$, are set by which EoS we chose for the NS. We note that $\beta < 0$ guarantees that $M_{max}$ decreases with increasing $P$. From this, one can derive the EoS critical rotation period $P_c$, above which the NS no longer has sufficient centrifugal force to oppose its own collapse \citep{Sun},
\begin{equation}
    P_c = \left(\frac{M - M_{\rm TOV}}{\alpha M_{\rm TOV}}\right)^{1/\beta}.
\end{equation}
For simplicity, we assume that the remnant initial rotation period is $P_0$ = 1~ms, which is a conservative assumption in the sense that it predicts fewer magnetars than assuming breakup rotation. Then, we consider that any remnant with a rotation period longer that its critical period collapses to a BH. 

It shall be noted that $M_{\rm TOV}$, the maximum mass of the nonrotating NS, is very uncertain. Thus, the outcome and evolution of the BNS merger remnant based on its mass is uncertain as well, all the more so considering $M_{\rm TOV}$ is estimated assuming no rotation and neglecting temperature, which are both inaccurate hypotheses to describe new-born magnetars. It is then interesting to consider various EoS and also important since no observational or theoretical consideration allows us to confidently favor any specific one at present. Our phenomenological approach to millisecond magnetar simulation allows us to simply compare different EoS. The list of considered EoS for this project has been taken from \cite{data_EoS} and is reported in Table \ref{tab:list_EoS}.

\begin{table*}[ht]
    \centering
    \caption{Considered main EoS characteristics \citep[source:][Table 1]{data_EoS}.}
     \label{tab:list_EoS}
    \begin{adjustbox}{max width=\textwidth}
    \begin{tabular}{c c c c c c}
    \hline\hline    
    Name & $M_{\rm TOV}(M_\odot)$ & $\alpha$ & $\beta$ & $I_{rot}$  (g~$\mathrm{cm}^{-2}$) & $R$ (km)\\ 
    \hline 
    Shen & 2.18 & $4.68\times10^{-10}$ & -2.738 & $4.68\times10^{45}$ & 12.40\\ [1ex]
    BSk21 & 2.28 & $2.81\times10^{-10}$ & -2.75 & $4.37\times10^{45}$ & 11.08\\ [1ex]
    GM1 & 2.37 & $1.58\times10^{-10}$ & -2.84 & $3.33\times10^{45}$ & 12.05\\ [1ex]
    DD2 & 2.42 & $1.37\times10^{-10}$ & -2.88 & $5.43\times10^{45}$ & 11.89\\ [1ex]
    DDME2 & 2.48 & $1.97\times10^{-10}$ & -2.84 & $5.85\times10^{45}$ & 12.09\\
    \hline
    \end{tabular}
    \end{adjustbox}
\end{table*}

\subsubsection{Examples of simulated light-curves}
\label{sect:light_curves}
\begin{figure*}
    \centering
    \begin{subfigure}{.33\textwidth}
        \centering
        \includegraphics[width=1.\linewidth]{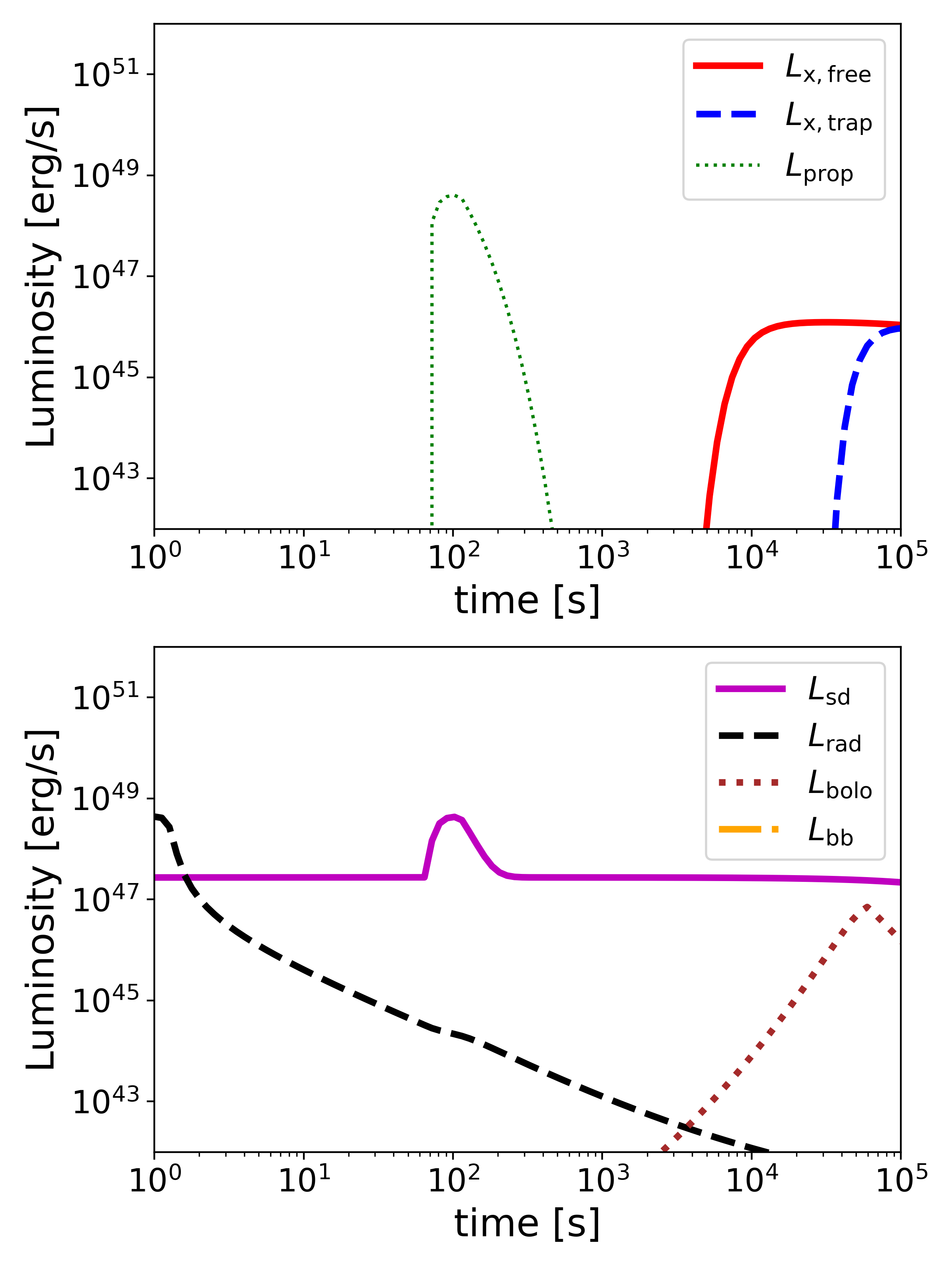}
        \caption{}
    \end{subfigure}
    \begin{subfigure}{.33\textwidth}
        \centering
        \includegraphics[width=1.\linewidth]{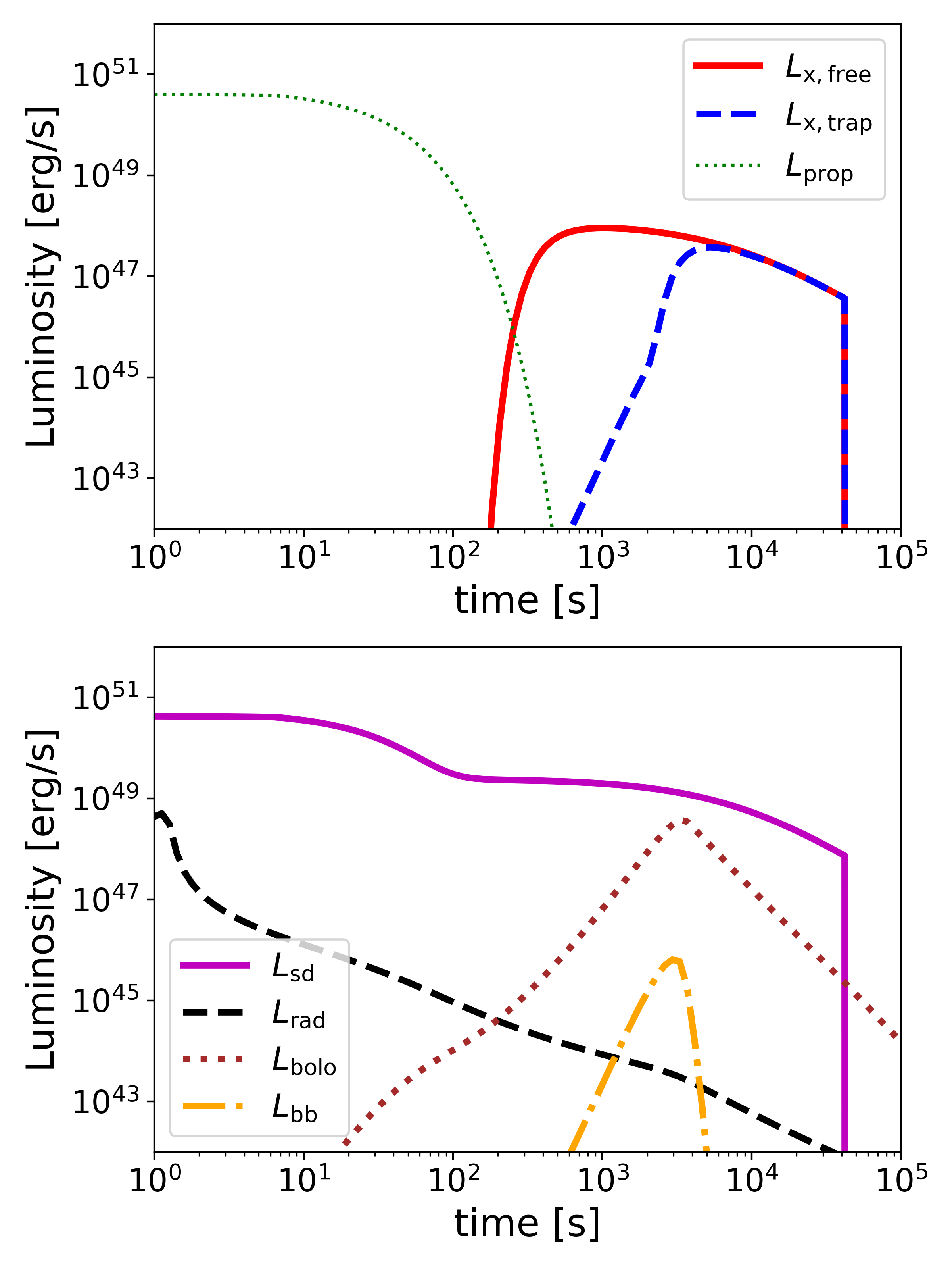}
        \caption{}
    \end{subfigure}%
    \begin{subfigure}{.33\textwidth}
        \centering
        \includegraphics[width=1.\linewidth]{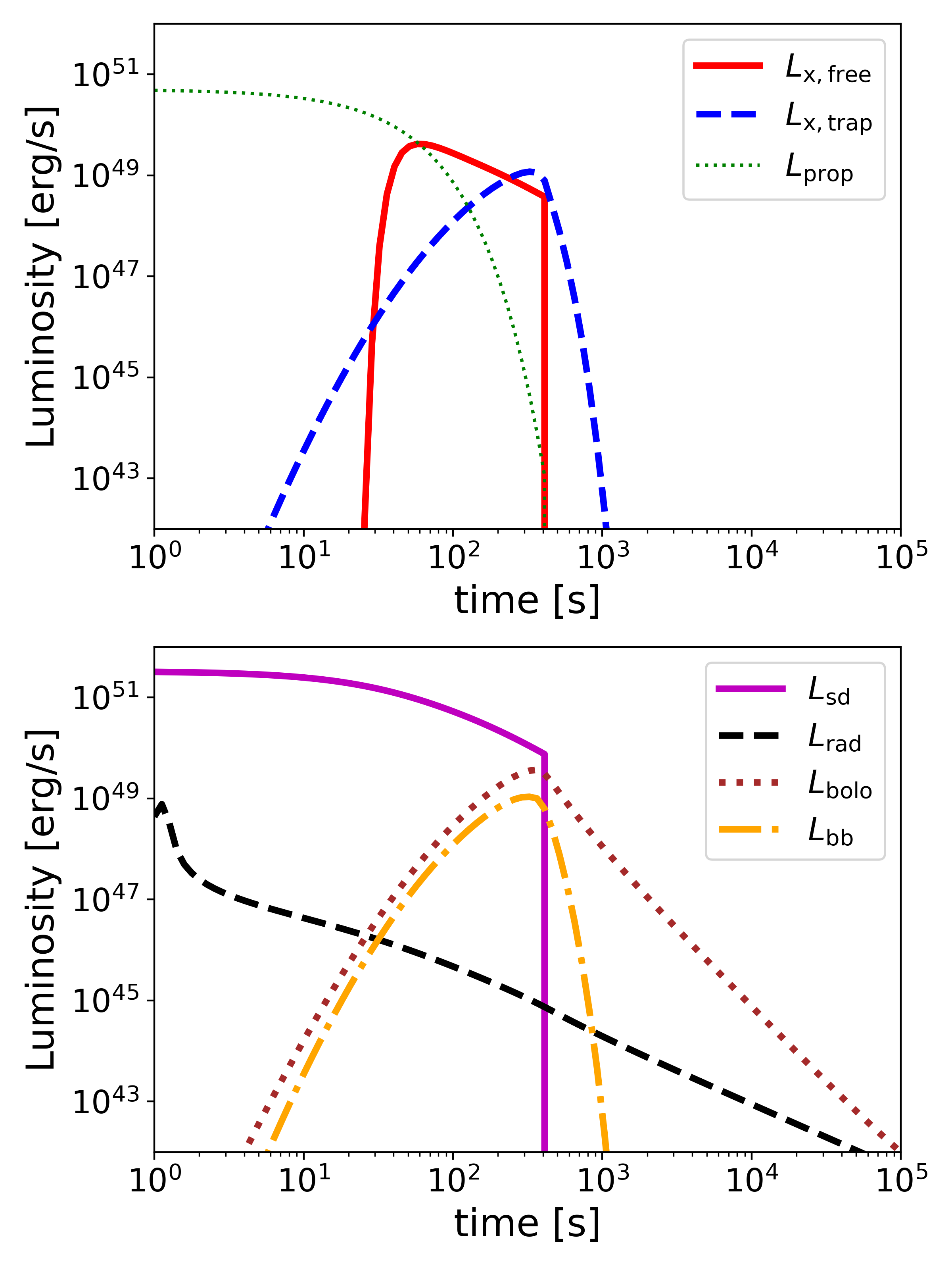}
        \caption{}
    \end{subfigure}
    \caption{Simulated light-curves for a magnetar with a dipolar magnetic field of 10$^{14}$~G (panel a), 10$^{15}$~G (panel b), and 10$^{16}$~G (panel c) assuming the DD2 EoS with a 2.37~M$_\odot$ NS mass. The top panels show the emission from the two considered viewing angles, and the bottom panels detail the physical origins of the emission. Times are measured relative from merger.}
    \label{grb_output}
\end{figure*}

The X-ray light-curve of an early X-ray signal from an NS remnant following a GW event is shown for three cases in Fig. \ref{grb_output}, where each panel represents the predicted light-curves differing only by the NS dipolar magnetic field (set at 10$^{14}$~G on the left, 10$^{15}$~G on the middle, and 10$^{16}$~G on the right), and every other parameter (NS mass, EoS, etc.) fixed otherwise.

In the top panel of Fig. \ref{grb_output} we plot the X-ray light-curves expected from the free zone and the trapped zone, in red and blue respectively. The magnetar X-ray luminosity increases sharply as the ejecta becomes optically thin and when the propeller regime starts, flattens when the dipolar luminosity takes over (e.g., late-time plateau), and drops when the magnetar collapses into a BH. In the bottom panel, the magnetar emission related to its spin-down ($L_{\text{sd}}$) is in magenta, and results of the contributions of the propeller ($L_{\text{prop}}$) luminosity, in green, and the dipolar luminosity ($L_{\text{dip}}$), not shown here. In all panels, we can see that the propeller regime (see, e.g., the bump in the spin-down luminosity in the bottom left panel) ends before the dynamical ejecta becomes optically thin, and thus contributes to the luminosity only through its effect of the NS angular velocity. 
The radioactive emission of the r-process ejecta ($L_{\text{rad}}$) is shown in black and is subdominant. Last, we show the X-ray blackbody emission ($L_{\text{bb}}$) in orange, and the bolometric blackbody luminosity ($L_{\text{bolo}}$) in brown (see \cite{energy_injection} for more detail). 

The sharp drop in luminosity displayed in the second and third columns of Fig. \ref{grb_output} reflects the collapse of the magnetar into a BH.
For the $10^{16}$~G dipolar magnetic field magnetar, increasing the magnetic field by an order of magnitude means a more efficient spin-down energy ejection. This entails a higher global luminosity (see, e.g., the spin-down and black-body luminosity), but also a faster braking of the magnetar, causing it to collapse into a BH much quicker, at around $t = 3\times 10^2$~s (about 5~min after the merger). For the $10^{15}$~G dipolar magnetic field magnetar, this collapse is predicted at around $t = 4 \times 10^4$~s (more than 1h30 after the merger). For the $10^{14}$~G dipolar magnetic field, the magnetar lives on longer than $10^5$~s.

\subsubsection{Observational constraints}
\label{sect:obs_constraints}

The last step of the EM simulation consists in obtaining instrumental capabilities, in particular from missions designed to hunt EM counterparts to GW such as the Space Variable Object Monitor (SVOM) \citep{SVOM,SVOM_2}, and compare them to predicted fluxes. Following SVOM observation strategy, the confident BNS GW detection triggers the Target of Opportunity (ToO) program with a high priority. The alert is then uplinked to the space platform via \href{http://en.beidou.gov.cn}{Beidou} \citep{beidou}, within 15~s to 120~s. This enables the satellite to slew in typically 2~min, and point its instruments in search for the EM counterpart.

Instruments with a narrower field of view (FoV) usually offer the best chances of detection for extragalactic magnetars considering their distance, and the fact that their luminosity (outside the jet) is of about 10$^{50}$ erg s$^{-1}$ or lower. With still relatively wide FoVs ($\sim$ 1~deg$^2$) to rapidly pave the large error boxes expected from a GW alert, MXT on board SVOM \citep{mxt} and FXT \citep{EP_FXT} aboard \textit{Einstein} Probe (EP; \citealt{EP}) are promising instruments for a near-future millisecond magnetar detection. The two instruments share the same energy band (0.2 $-$ 10~keV).

A significant delay between the GW alert and the beginning of observations of the EM counterpart is expected, making the detection of the millisecond magnetar time sensitive. This is due to two reasons, intrinsic to the GW detection : the processing time of the alert, and the low localization accuracy. For the former, we consider that the reconstructed sky localization using Bayesian inference may be released under a conservative $\sim$ 1~h \citep{Chaudhary_2024}. Then, the start of observations of the EM counterpart largely depends on the GW localization accuracy. We set our possible beginning of X-ray observations at $\sim$ 1~h after the merger, which corresponds to the case of a detection with a $\Delta \Omega \approx 1deg^2$ localization accuracy, or a lucky find in one of the instrument first tiles. 

To cover all GW detection scenarios, a compromise has to be found between the X-ray instrument tiling speed and exposure time. For MXT, the strategy is to divide an orbit into five exposures, of almost 10 minutes each. In this way, most of the error box for a relatively well-localized O5 GW detection with a $\sim$ 30~deg$^2$ localization accuracy will be observed in six orbits, which represents a duration of nine hours after coalescence. Now, with the knowledge of the exposure time, it remains to examine the corresponding MXT sensitivity.
With a 10 000~s exposure, MXT sensitivity at 5 $\sigma$ between 0.3~keV and 6~keV is of $\sim$10$^{-12}$~erg~cm$^{-2}$~s$^{-1}$.
The photon noise, which is the limiting factor to the flux sensitivity, evolves as $\sqrt{t}$ following a Poisson statistics.
Taking the nominal 9-minute exposure, the detection threshold flux for MXT is then of the order of $F^{MXT}_{thres}(t_{exp} = 9\ \mathrm{min}) = 4.30 \times 10^{-12}\ \mathrm{erg\ cm^{-2}\ s^{-1}}$.

The same computation operated on FXT yields a detection threshold flux one order of magnitude lower, of $F^{FXT}_{thres}(t_{exp} = 9min) = 4.30 \times 10^{-13}\ \mathrm{erg\ cm^{-2}\ s^{-1}}$ \citep{EP_FXT}.
However, EP observation strategy in case of a GW alert is first to cover the GW errorbox with the Wide FoV X-ray Telescope WXT, with a 3800~deg$^2$ FoV \citep[see][]{EP}, with 2400~s tiles. The sensitivity of WXT at this stage will thus be $1.68 \times 10^{-11}$~erg~cm$^{-2}$~s$^{-1}$.
Then, FXT will look for the EM counterpart, with 5~min exposures, corresponding to a flux sensitivity of $5.88 \times 10^{-13}$~erg~cm$^{-2}$~s$^{-1}$. We compared the predicted millisecond magnetar fluxes to the instrument sensitivity with a 9~min exposure for each in order to compare the two instruments under the same observation conditions.

To this day, the X-ray instrument(s) that will operate at the same time as the next generation of GW interferometer is still unclear. If their performance is somewhat maintained, FXT and MXT may still be considered. SXI and XGIS aboard THESEUS \citep{theseus_1, theseus_2}, with a planned launch date (2037) near ET start of operations ($\sim $ 2035), as well as WFI and X-IFU aboard Athena \cite{athena}, provisionally due for launch in the early 2030s, will probably be the instruments to consider for this epoch.

Finally, an upper boundary on the expected fluxes can be discussed based on MAXI observations. MAXI \citep[Monitor of All-sky X-ray Image;][]{maxi} is a telescope collecting X-ray data on board the ISS since 2009. Over its entire effective period, MAXI has detected sources with a typical flux limit (2-10 keV) of $\sim 10^{-8}$~erg~cm$^{-2}$~s$^{-1}$. 
Transients with fluxes above this value, and occurring with a rate of $\sim$10$^{-1}$~yr$^{-1}$ or above, potentially originating from nearby magnetar-powered X-ray events, can already be considered unlikely, as they would have been detected in the past as extra-galactic X-ray transients of unknown origin.

\section{Predictions of a millisecond magnetar detection}
\label{sect:predictions}
We first study detection prospects for fiducial parameters in Sect. \ref{sect:det_prospects}, and then discuss their dependence on the magnetar parameters in Sect. \ref{sect:magnetar_param}.
Our foreseen number of millisecond magnetar detections is given in Sect. \ref{sect:number_det}, and we discuss the consequence of our results for an observation strategy in Sect. \ref{sect:obs_strategy_magnetar}.

\subsection{Millisecond magnetar detection prospects}
\label{sect:det_prospects}
To focus on BNS population-wide effects, we set the millisecond magnetar intrinsic parameter values at $B = $ 10$^{15}$~G and $P_0 = $ 1~ms throughout this section.

\subsubsection{Free-zone versus trapped-zone detection}
\label{sect:freevstrapped_results}
As discussed in Sect. \ref{sect:obs_geom}, the observer viewing angle impacts the millisecond magnetar EM detectability, to an extent dependent on the GW detector configuration. Here, we show the EM detectability at the typical satellite slew time following a GW alert of one hour after the merger. 

\begin{figure*}
    \centering
    \begin{subfigure}{.45\textwidth}
        \centering
        \includegraphics[width=1.\linewidth]{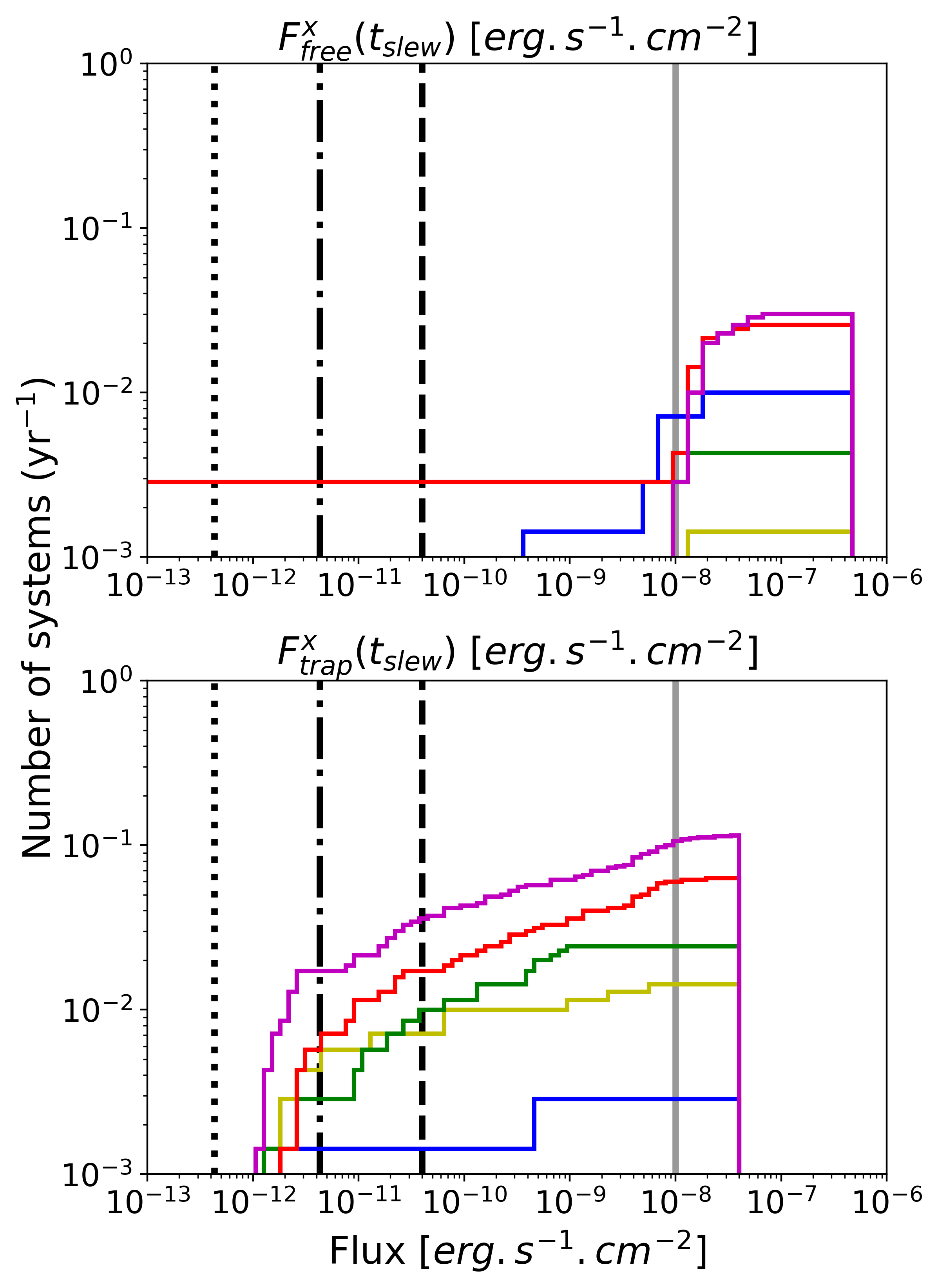}
        \caption{}
    \end{subfigure}%
    \begin{subfigure}{.45\textwidth}
        \centering
        \includegraphics[width=1.\linewidth]{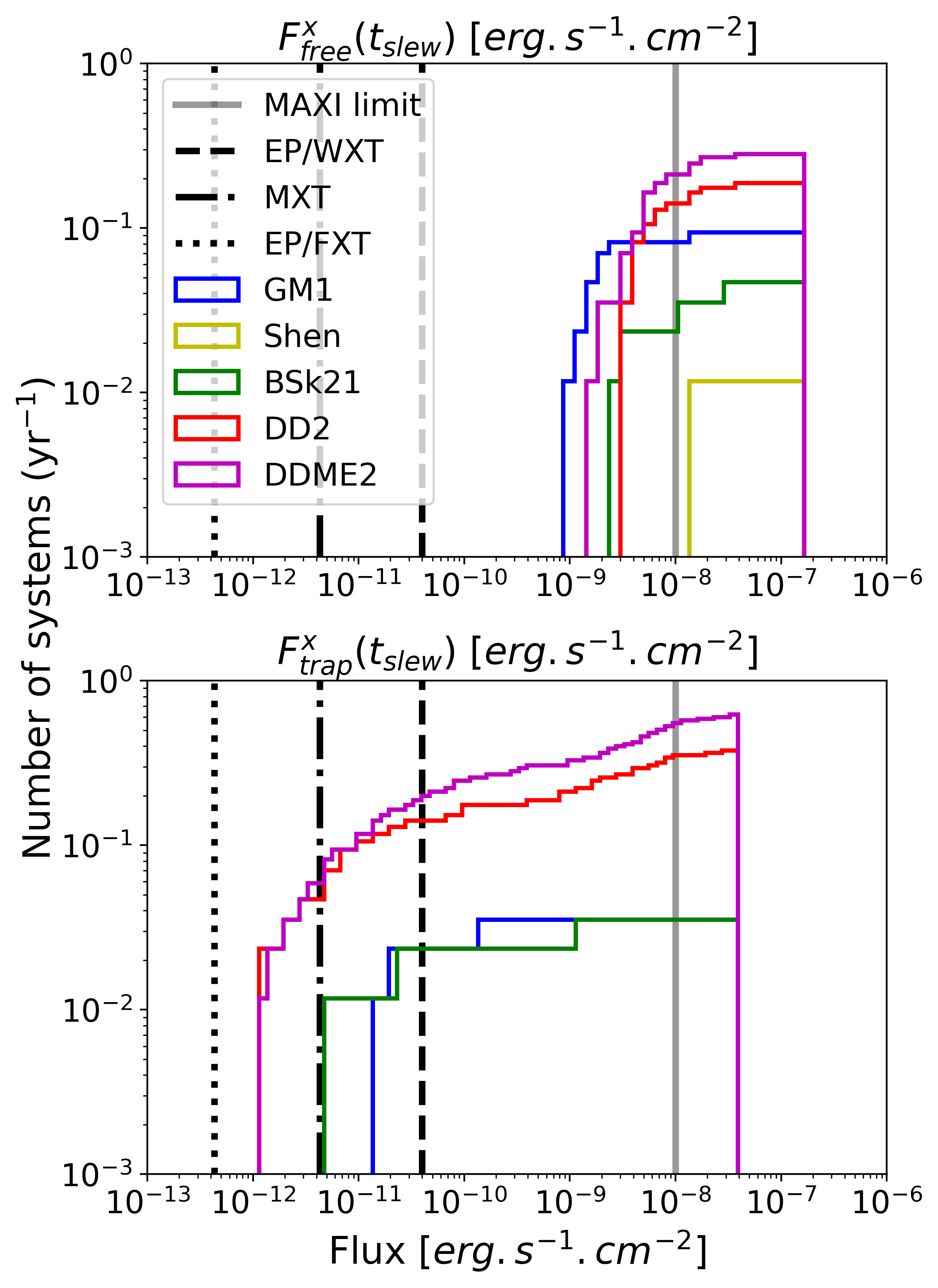}
        \caption{}
    \end{subfigure}
    \caption{Cumulative histograms showing the repartition of X-ray flux in the MXT band 1~h after the merger for GW well-localized events. The curves in different colors correspond to the five different EoS we considered. The vertical black lines indicate X-ray instruments sensitivity threshold for SVOM/MXT (dash-dotted), EP/WXT (dashed) and EP/FXT (dotted). The left and right panels correspond to O4 and O5 LVK(I) runs, respectively. The top and bottom panels correspond to the flux in the free and trapped zone, respectively. The shaded line represents the MAXI nondetection limit: fluxes above this limit and with an occurrence rate of $\sim$10$^{-1}$~yr$^{-1}$ or above are likely over-estimated.
    } 
    \label{freevstrap_LVK(I)}
\end{figure*}

For LVK(I), Fig. \ref{freevstrap_LVK(I)} shows cumulative histograms of the X-ray flux for our selection of EoS, and compares it to relevant instrument detection threshold. 
The figure is separated into two panels, with the top panel corresponding to systems detected in a free zone configuration, and the bottom panel a trapped zone configuration. 
For O4, the number of systems detected in GW at larger distances is limited. Millisecond magnetars within this range would be detectable in the free and trapped zone configurations, with a slight preference for detection in the trapped zone due to the longer emission duration. This is illustrated in the left panel of Fig. \ref{freevstrap_LVK(I)}, where the cumulative histograms show a higher number of detected events in the trapped zone than in the free zone, reaching up to 1 $\times$ 10$^{-1}$ detections per year compared to 2 $\times$ 10$^{-2}$ detection per year, respectively, assuming the DDME2 EoS. Due to its larger BNS range, the O5 configuration indicates more balance breakdown of the expected flux in the free and trapped zones (see the right panels of Fig. \ref{freevstrap_LVK(I)}). 
Due to the moderate occurrence (10$^{-1}$~yr$^{-1}$ max., consistently with the lack of detection so far) of joint X-rays and GW detections in O4, our predicted fluxes are compatible with MAXI nondetections. 
For the two GW configurations, we note that emission of the millisecond magnetar in the free zone configuration is expected at high fluxes. Thus, with a mean free zone flux at $\sim$ 2 $\times$ 10$^{-8}$~erg~cm$^{-2}$~s$^{-1}$ for O4, and $\sim$ 1 $\times$ 10$^{-8}$~erg~cm$^{-2}$~s$^{-1}$ for O5, we find millisecond magnetars in the free zone to be detectable by WXT aboard EP, although no EP transient related to a millisecond magnetar has been reported in the literature so far.
By contrast, EP would miss up to 20\% of magnetars in the trapped zone. The moderately more probable detection in this zone shall be operated by MXT-like pointed instruments. This statement is at the heart of our proposed observed strategy, developed in Sect. \ref{sect:obs_strategy_magnetar}.

\begin{figure*}
    \centering
    \begin{subfigure}{.45\textwidth}
        \centering
        \includegraphics[width=1.\linewidth]{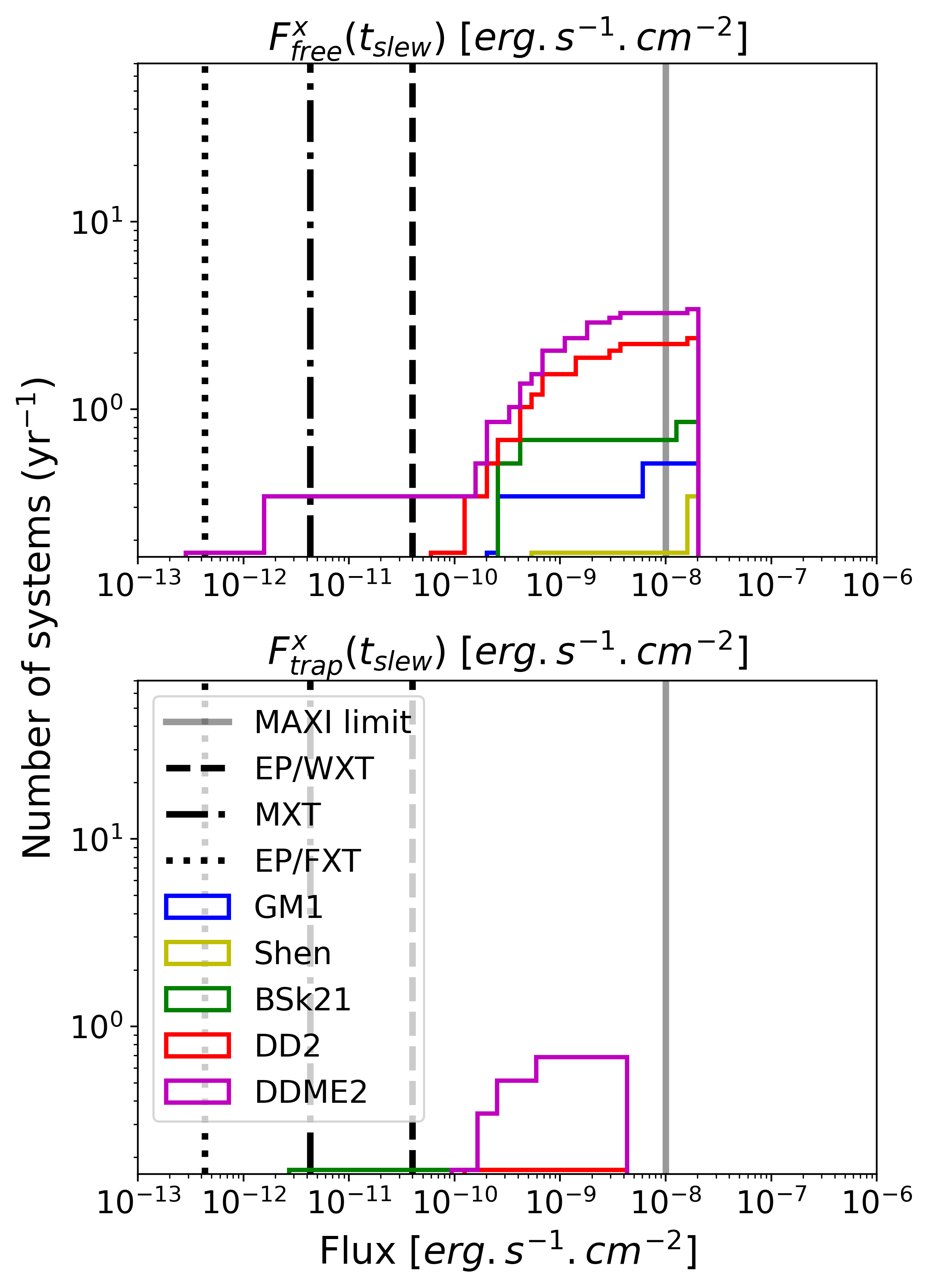}
        \caption{}
    \end{subfigure}%
    \begin{subfigure}{.45\textwidth}
        \centering
        \includegraphics[width=1.\linewidth]{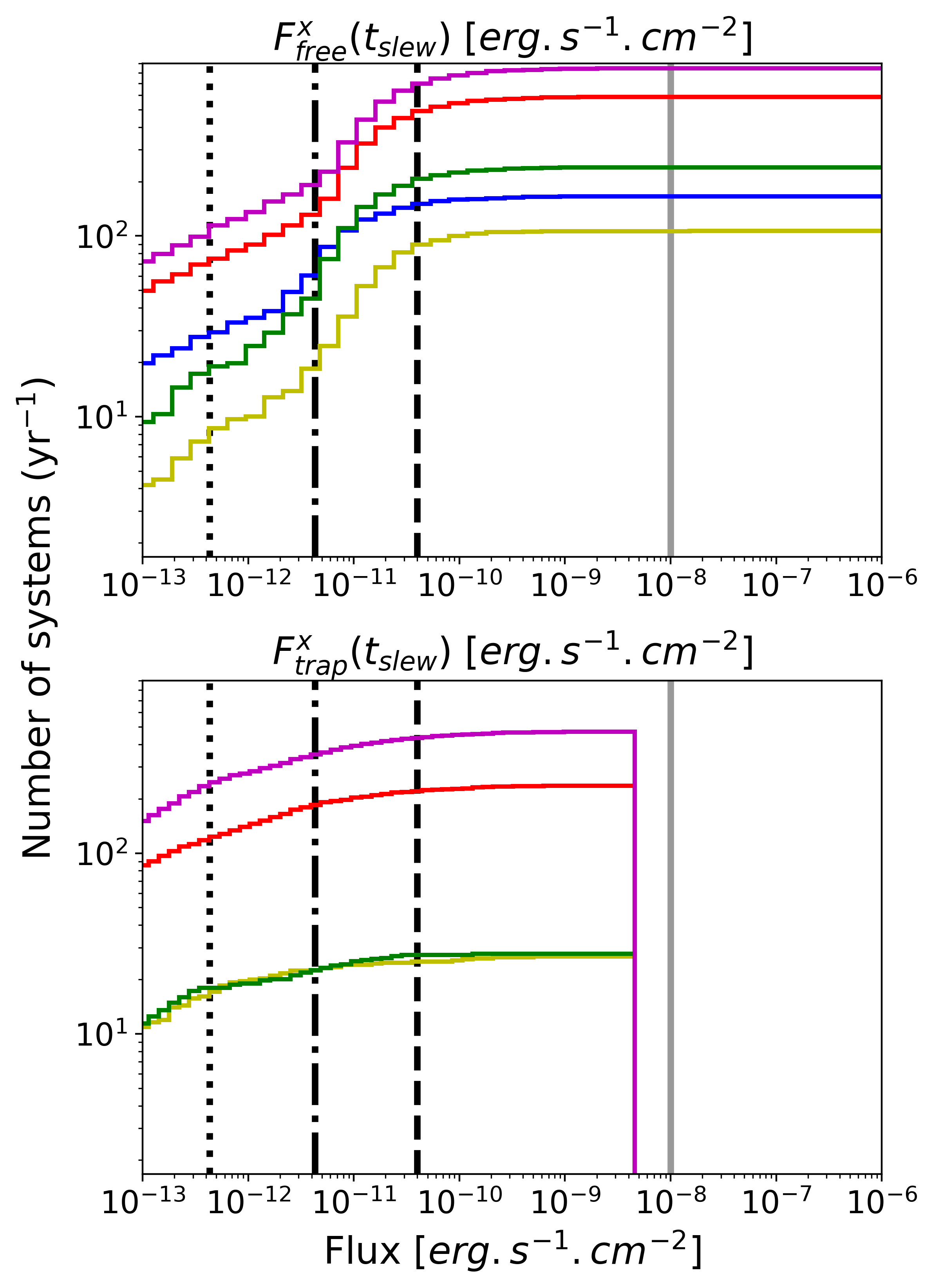}
        \caption{}
    \end{subfigure}
    \caption{Same as Fig. \ref{freevstrap_LVK(I)}, but for ET $\Delta$ 10~km (left panels) and ET $\Delta$ 10~km + 1 CE with 40~km arms (right panels; note the different scale). }
    \label{freevstrap_ET_LVK(I)}
\end{figure*}

For the next generation of GW interferometers, the same figure is pictured in Fig. \ref{freevstrap_ET_LVK(I)}. The larger distance reach implies fainter fluxes. Consequently, most events fall below the detection threshold of MAXI. This also means that a significant portion of expected fluxes will fall below the detection threshold of even the most sensitive small FoV X-ray instruments. The results vary significantly depending on the considered ET geometry. Indeed, the largest distance within reach for ET in a triangle of 10~km arms, but moderate localization compared to ET with CE, implies a favored detection for more distant and luminous systems, i.e., in the free zone. For ET with CE, we state a mild favoring of the free zone detection.

\subsubsection{Flux evolution over time}
\label{sect:det_with_time}
To fully evaluate the typical millisecond magnetar detectability, it is also important to consider the evolution of its luminosity with time. For this, we only consider the trapped zone. Indeed, the emission from the free zone, being less reprocessed than in the trapped zone, rarely outlives the millisecond magnetar lifetime. The emission duration in the free zone is thus limited to the magnetar time of collapse (if any), typically ending before our predicted start of observations for more than half of the magnetars with a magnetic field of 10$^{15}$~G or above (see, e.g., Fig. \ref{hist_time_col}). Thus, we estimated the fluxes from the potential magnetars in the trapped zone at fixed times after the merger to monitor the evolution from SMNS merger remnant on top of stable NS.

The detailed evolution of flux with time is illustrated in Fig. \ref{fluxvstime_ETtriangle} for ET in a triangle configuration, and is reported in Appendix \ref{ap:flux_vs_time} for ET with CE (Fig. \ref{fluxvstime_ET_CE}), O4 (Fig. \ref{fluxvstime_O4}), and O5 (Fig. \ref{fluxvstime_O5}). For ET, since the systems are detected at larger distances, the effect of the cosmological redshift significantly impacts the flux evolution with time. In particular, the time dilation delays the time of maximum detectability by one to two hours compared to LVK(I) observing runs. This is verified in Fig. \ref{detvstime}, which shows the detectability of the magnetar with time depending on the configuration (GW detector and viewing angle).
As seen on this figure, for ET, we find a maximum detectability of the millisecond magnetar between three to four hours after the merger depending on the observer viewing angle, in particular for the trapped zone, in which configuration $\sim$47\% of all detections are expected at four hours after the merger, compared to $\sim$40\% at three hours after the merger.
\begin{figure*}
    \centering
    \includegraphics[width=.94\linewidth]{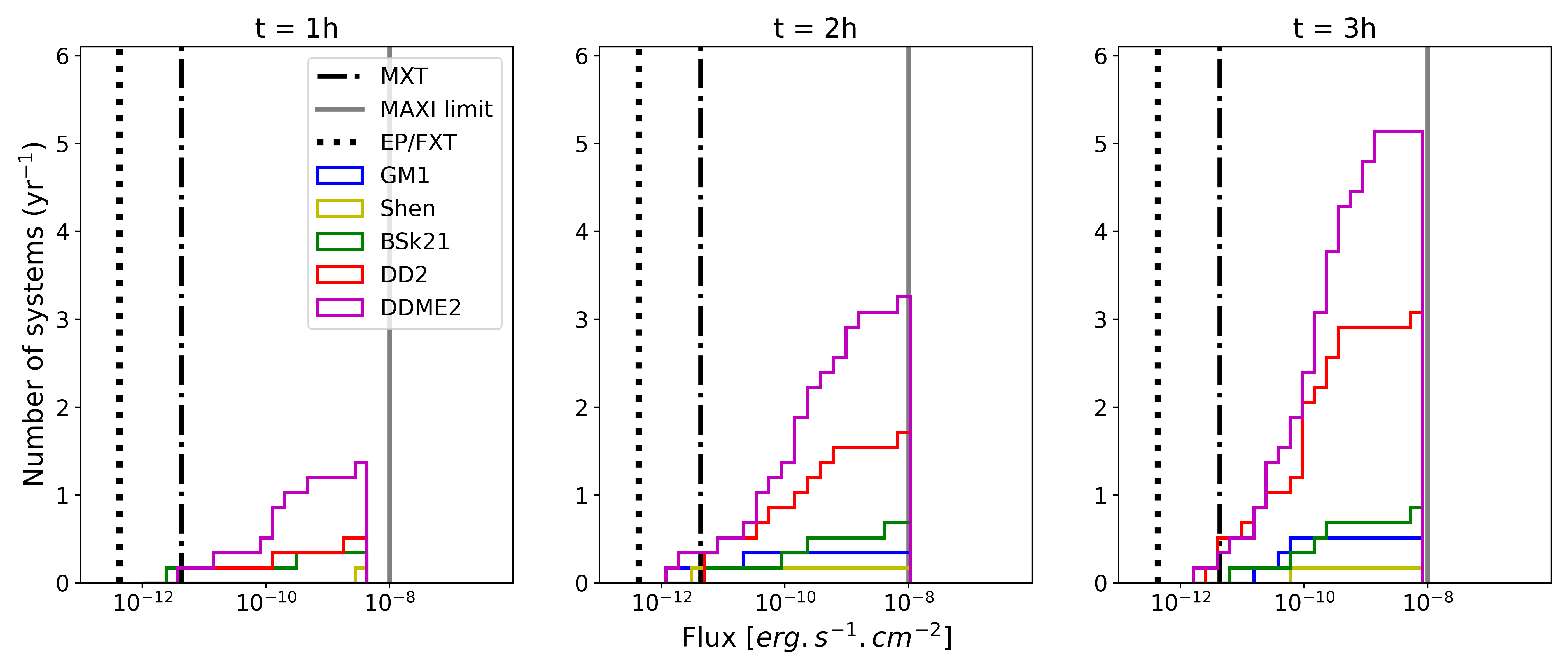}
    \caption{Cumulative histograms of the X-ray flux in the trapped zone for ET $\Delta$ 10~km. Different panels correspond to different times after merger: 1~h (left), 2~h (middle) and 3~h (right). Vertical lines are the same features as previous figures. }
    \label{fluxvstime_ETtriangle}
\end{figure*}

For LVK(I), the emission of the trapped zone peaks around two hours after the merger, as can be seen for O5 in Fig. \ref{detvstime}, reaching 60\% detectability at this time. This indicates that the mean time at which the dynamical ejecta becomes optically thin (cosmological time dilation aside) is close to two hours (see also Fig. \ref{grb_output}). However, this behavior of the X-ray trapped zone luminosity varies depending on the EoS. For instance, Bsk21 \citep{eos_bsk21} predicts millisecond magnetars whose emission typically peaks at one hour after the merger. GM1 \citep{eos_gm1} appears as an outlier among all EoSs, predicting emission peaking at later times (i.e., three hours after the merger) than other EoSs. Our interpretation for this particular EoS is that this is the expression of its lower moment of inertia, evaluated by \cite{data_EoS} to be lower than any other of our EoSs moment of inertia by at least 30\% (see Table \ref{tab:list_EoS}). Indeed, a lower moment of inertia fastens the collapse, while the relatively high $M_{\rm TOV}$ (2.37~M$_\odot$) favors the formation of stable NS, which deposit less kinetic energy into the ejecta, explaining the predicted fluxes at later times only.

\begin{figure}
    \centering
    \includegraphics[width=.9\linewidth]{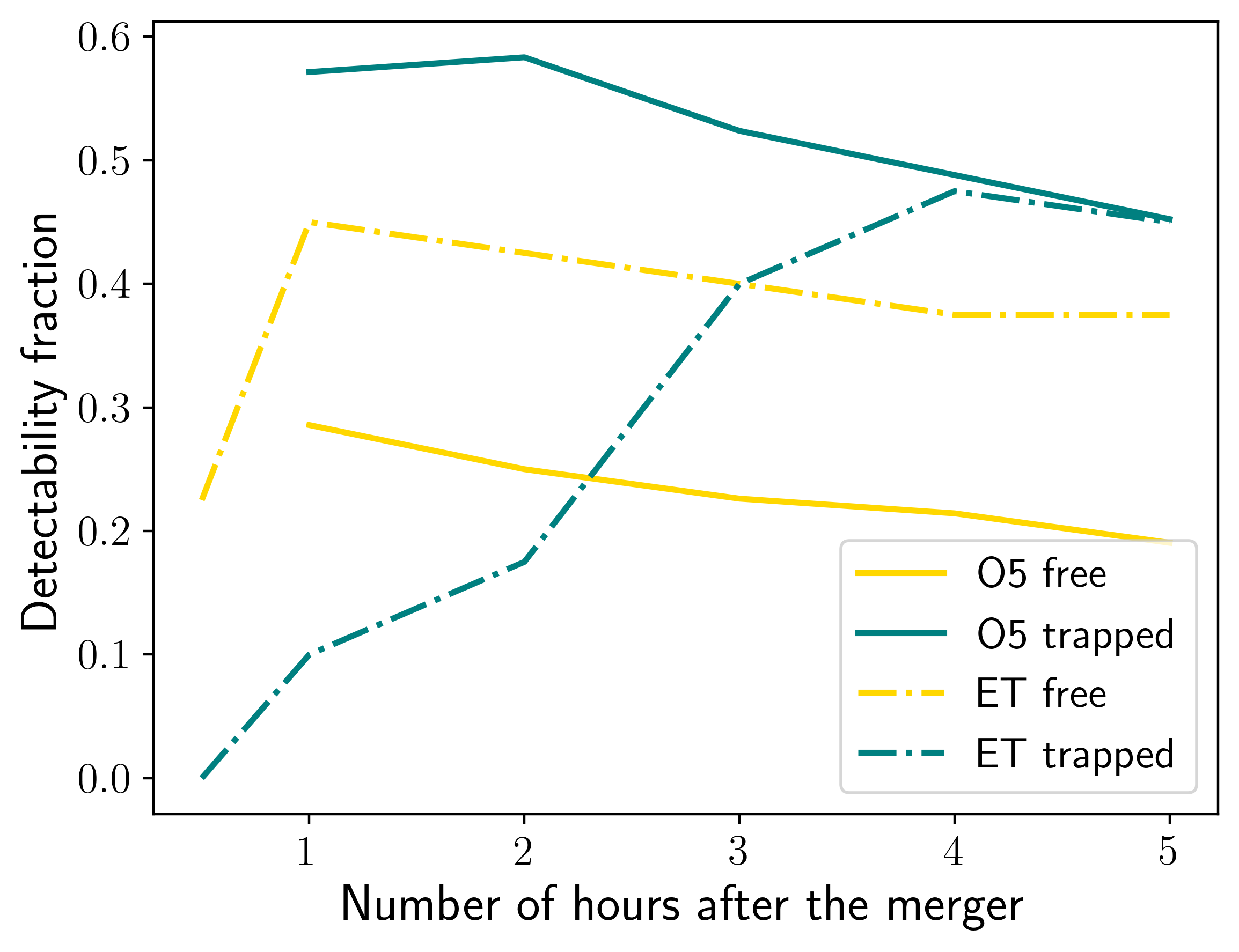}
    \caption{Detectability of a formed 10$^{15}$~G millisecond magnetar as a function of time. Results are shown for the current (full) and next generation (dashed) of GW interferometers, and distributed in the free (yellow) and trapped (blue) zones configurations, assuming the DD2 EoS and MXT X-ray sensitivity. The number of detections at a given time is normalized by the total number of magnetar detections over all times and configurations for the considered GW configuration. }
    \label{detvstime}
\end{figure}

\subsection{Effect of the magnetar characteristics}
\label{sect:magnetar_param}
For fiducial magnetar parameters, we provide a comprehensive study of the influence of the magnetar characteristics on the results. 

\subsubsection{Collapse time}
\begin{figure}
    \centering
    \includegraphics[scale = 0.55]{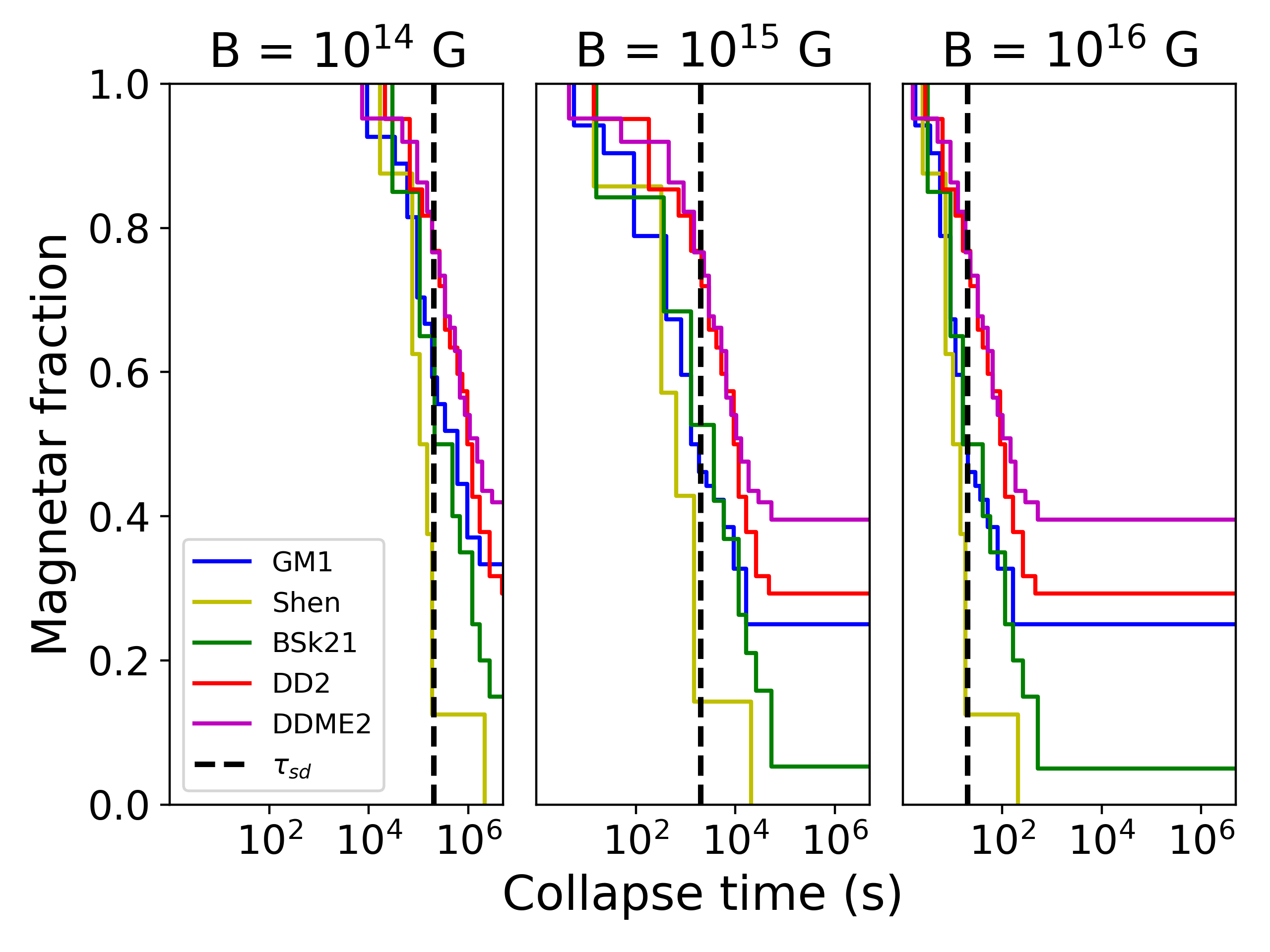}
    \caption{Surviving fraction of our magnetar population as a function of collapse time. Results are shown for three magnetar dipolar magnetic field values: $B = $ 10$^{14}$~G (left panel), $B = $ 10$^{15}$~G (middle panel), and $B = $ 10$^{16}$~G (right panel). The characteristic spin-down time $\tau_{sd}$ for each magnetic field is indicated by the dashed vertical line. The different colors correspond to different EoS.} 
    \label{hist_time_col}
\end{figure}

As discussed previously, one of the main reason why the detectability of the millisecond magnetar is time-sensitive is due to the rapid collapse of the magnetar into a BH. It is thus interesting to look at the predicted time of collapse (if any) on our newly formed magnetar population, depending on the EoS.
The results shown so far are based on predictions with magnetars with a fixed dipolar magnetic field $B = $ 10$^{15}\,\mathrm{G}$. In Fig. \ref{hist_time_col} we show the distribution of collapse time for three different magnetic field simulations.  

Assuming that the spin-down of the magnetar is dominated by dipole emission, the characteristic spin-down time can be estimated from the dipole luminosity (Eq. \ref{L_dip}) as follows \citep{energy_injection_2, Sun}:
\begin{equation}
\label{eq:time_sd}
    \tau_{sd} = \frac{3c^3I}{B^2R^6\Omega_0^2}.
\end{equation}
This indicates that an increase of dipolar magnetic field of one order of magnitude induces a spin-down time shortened by a factor of 100, as is visible in Fig. \ref{hist_time_col}, and as can also be stated in Fig. \ref{grb_output}. Quantitatively, the characteristic spin-down times are 
$2.0 \times 10^{5}\,\mathrm{s}$, 
$2.0 \times 10^{3}\,\mathrm{s}$, and 
$2.0 \times 10^{1}\,\mathrm{s}$ 
for magnetic field strengths of 
$10^{14}\,\mathrm{G}$, $10^{15}\,\mathrm{G}$, and $10^{16}\,\mathrm{G}$, respectively.

The simulation at $B =$ 10$^{16}$~G shows that all SMNS at this magnetic field collapse before 10$^3$~s, with only stable NS surviving after that time. This excludes the possibility of a high magnetic field magnetar as a central engine of a plateau longer than $\sim$ 30~min for the lower $M_{\rm TOV}$ EoSs.

Finally, regarding the impact of the magnetic field on the detection, the left panel of Fig. \ref{hist_time_col} reveals that the absence of flux from one hour after the merger does not necessarily mean that no flux will be detected later. 
It is particularly the case of lower magnetic field magnetar: for a $B =$~10$^{14}$~G magnetar, the magnetar lifetime can significantly exceed one hour (Fig. \ref{grb_output}). 
In addition, if the lower magnetic field is in a trapped configuration, our modeling work presented in Sect. \ref{sect:grb.py} reveals that since less kinetic energy is injected in the ejecta, it becomes optically thin later, indicating that if present, the millisecond magnetar might be exclusively observable hours after the merger.

\subsubsection{Dipolar magnetic field and magnetar detectability}
\begin{figure}
    \centering
    \includegraphics[scale = 0.7]{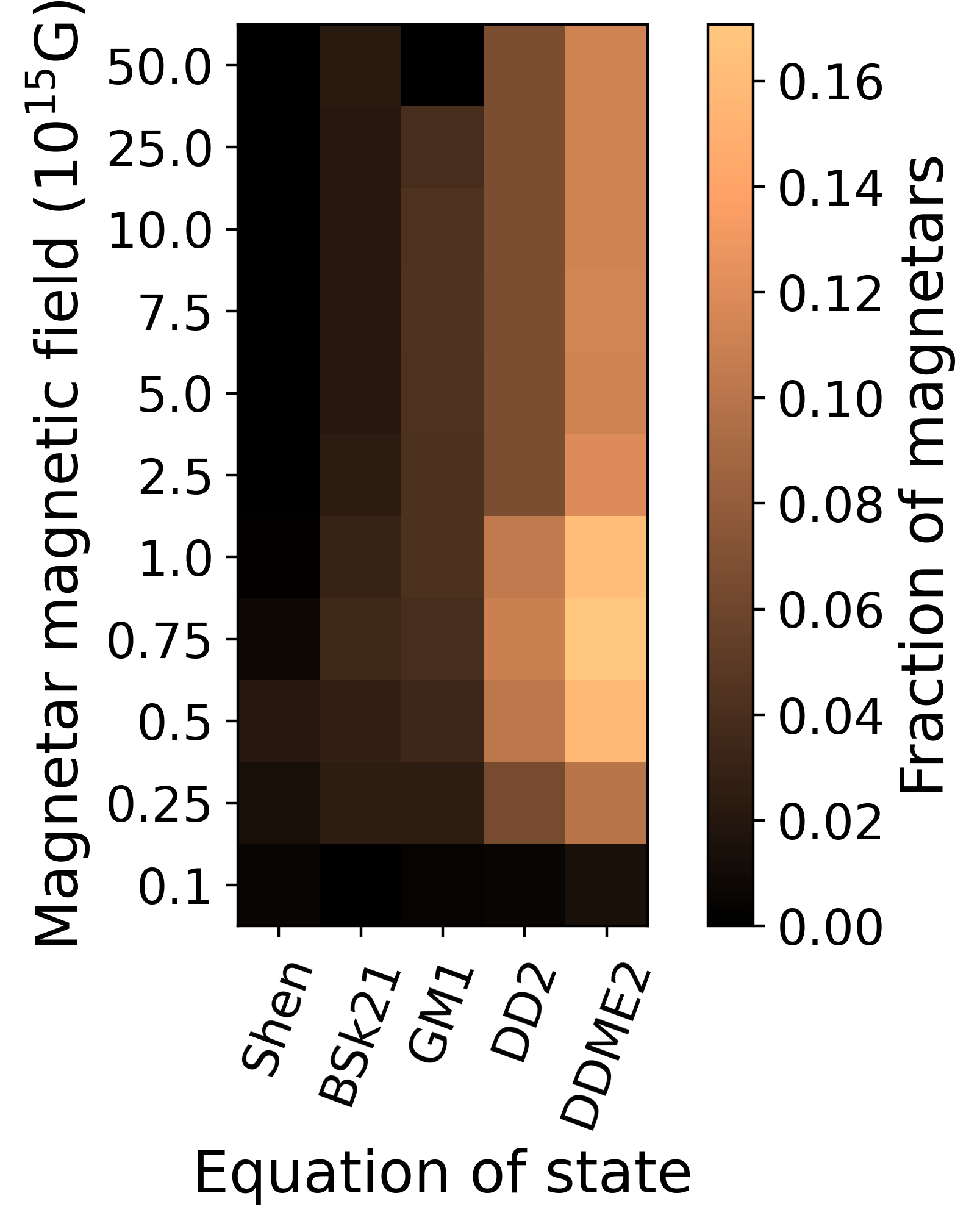}
    \caption{Fraction of well-localized GW events for which magnetar emission is detectable by MXT 1~h after merger, shown as a function of the EoS and dipolar magnetic field, obtained from LVKI O5 simulations. }
    \label{param_exploration}
\end{figure}

To complete this study, we explore the impact of the magnetar magnetic field on the fraction of detectable magnetars. In Fig. \ref{param_exploration}, each pixel quantifies by its colors the fraction of magnetars with a predicted flux at one hour after the merger above MXT threshold sensitivity for a given magnetic field - EoS configuration. 
As hinted before, despite their longer lifetime, lower magnetic field magnetars are not the most easily detectable, since their injected energy often falls below the MXT threshold sensitivity. The maximum detectability at a given time results from a compromise between the energy injection rate and the millisecond magnetar lifetime, which at one hour after the merger occurs for 7.5$\times$10$^{14}$~G magnetars with a 16\% detectability among all well-localized GW events. Again, of course, the higher the predicted $M_{\rm TOV}$, the higher the chances of a millisecond magnetar formation and detection. 

\subsection{Number of detections}
\label{sect:number_det}
Finally, we provide the number of expected detections from our simulation results. 
For this, we assume a typical millisecond magnetar of $B = $ 10$^{15}$~G and $P_0 = $ 1~ms. These results have been obtained by evaluating the magnetar detectability at fixed times (in terms of hours after the merger), assuming that pointed instruments can begin observations one hour after the merger at the earliest. More precisely, we define a detection by the excess of flux compared to MXT detection threshold over all our considered times after the merger, i.e., between one hour (30 minutes for the next generation of GW interferometers for possible pre-merger detections) and five hours after the merger. We find that all millisecond magnetars are detectable before four hours after the merger for LVK(I) and before five hours after the merger for the next generation of GW interferometers (and $>$ 90\% are detected before three hours after merger for LVK(I) and before four hours after the merger for the next generation of GW interferometers). Though it may happen that some systems become detectable in between our considered fixed times, these systems probably stay detectable over sufficiently long time periods such that they are accounted for in our detection rates.

In Table \ref{tab:det_rates} we first give the number of well-localized GW events depending on the GW detectors, and we then consider the identification of a millisecond magnetar in the EM counterpart. The number of well-localized GW detections gives an idea of the size of the overall BNS population from which finding an EM counterpart, and thus millisecond magnetar smoking guns, is feasible. To estimate the statistical uncertainty for all of our detection rates, we used a nonparametric bootstrap method. Starting from the set of simulated GW events that are well localized, each event was labeled based on whether it led to a detectable millisecond magnetar. These detections were treated as Bernoulli trials (success or failure). We then generated 5000 bootstrap re-samplings of the original sample by drawing with replacement, computing the fraction of detectable magnetars in each resampled dataset. This produced an empirical distribution of detection fractions, from which we derived the confidence interval. The final uncertainty on the detection rate was obtained by scaling these fractions by the assumed rate of well-localized BNS events. The 3$\sigma$ confidence interval (corresponding to the 0.15$^{th}$ and 99.85$^{th}$ percentiles of the bootstrap distribution) was then reported as the statistical uncertainty on the detection rate.
We note that this uncertainty is only statistical, and does not account for the uncertainty due to the imprecise merger rate. In other words, this uncertainty quantifies the errors extracted from our obtained distributions, assuming the BNS merger rate $R = $ 100 Gpc$^{-3}$ yr$^{-1}$ to be accurate. 

Concerning the nature of the remnant, we find that between 85\% (DDME2) and 98\% (Shen) of BNS merger form a BH promptly, undergoing a hypermassive NS phase or not. Between 0\% to 6\% form stable NSs, and 2\% to 9\% survive as a supramassive NS. In total, between 2\% to 16\% of remnants may form long-lived millisecond magnetars.

The number of detections highly depends on the considered EoS. Thus, we define the lowest detection rate as corresponding to our results obtained using the EoS with the lowest $M_{\rm TOV}$ (Shen, $M_{\rm TOV} =$ 2.18 M$_\odot$), while the high rate corresponds to the highest $M_{\rm TOV}$ of our considered EoS (DDME2, $M_{\rm TOV} =$ 2.48 M$_\odot$). The results for each GW detector configuration are summarized in Table \ref{tab:det_rates}.

\begin{table*}[ht]
\caption{Detection rate of GW events with a localization accuracy $\le$ 50~deg$^2$ (first column) and of the millisecond magnetar (i.e., GW + EM) depending on the GW detector configuration. } 
\label{tab:det_rates}
\begin{center}
    \begin{tabular}{c c c c}
   \hline\hline
    \rule[-1ex]{0pt}{3.5ex} & GW detection rate (yr$^{-1}$) & Low $M_{\rm TOV}$ rate (yr$^{-1}$) & High $M_{\rm TOV}$ rate (yr$^{-1}$) \\ 
    \hline
    \rule[-1ex]{0pt}{3.5ex} O4 & 1.7$^{+0.01}_{-0.01}$ & 0.01$^{+0.03}_{-0.01}$ & 0.15$^{+0.04}_{-0.04}$ \\
    \hline
    \rule[-1ex]{0pt}{3.5ex} O5 & 13.53$^{+0.95}_{-0.94}$ & 0.01$^{+0.04}_{-0.01}$ & 0.99$^{+0.31}_{-0.30}$ \\ 
    \hline
    \rule[-1ex]{0pt}{3.5ex} ET $\Delta$ 10~km & 42.80$^{+5.94}_{-5.48}$ & 0.34$^{+0.86}_{-0.34}$ & 6.85$^{+3.26}_{-2.74}$ \\
    \rule[-1ex]{0pt}{3.5ex} ET $\Delta$ 10~km + 1 CE 40~km & 21769.28 $^{+146.72}_{-131.96}$ & 64.58$^{+7.83}_{-6.92}$ & 751.66$^{+24.91}_{-24.66}$ \\ 
   \hline
    \end{tabular}
    \end{center}
\tablefoot{These results were obtained assuming a 10$^{15}$~G magnetar, and showing the prediction for the lowest $M_{\rm TOV}$ EoS (Shen, $M_{\rm TOV}$~=~2.18 M$_\odot$; second column) and our highest $M_{\rm TOV}$ EoS (DDME2, $M_{\rm TOV}$=~2.48 M$_\odot$; third column). The statistical uncertainties were obtained using the bootstrap method.  }
\end{table*}

If the NS EoS is soft (i.e., has a low $M_{\rm TOV}$), no detections are expected with current GW interferometers. In this case, the first millisecond magnetar detection may become possible in the time of ET, with a predicted rate of 0.3$^{+0.9}_{-0.3}$~yr$^{-1}$ for ET in a triangle configuration, and is particularly probable in a configuration that favors a precise localization, bringing the predicted detection rate up to $\sim$ 65~yr$^{-1}$ with ET and CE. If the EoS is stiff (i.e., has a high $M_{\rm TOV}$) however, a first detection is possible with the current LVK(I) network, particularly in its O5 configuration with a 1.0$^{+0.3}_{-0.3}$~yr$^{-1}$ predicted detection rate. 

\begin{figure}
    \centering
    \includegraphics[scale = 0.5]{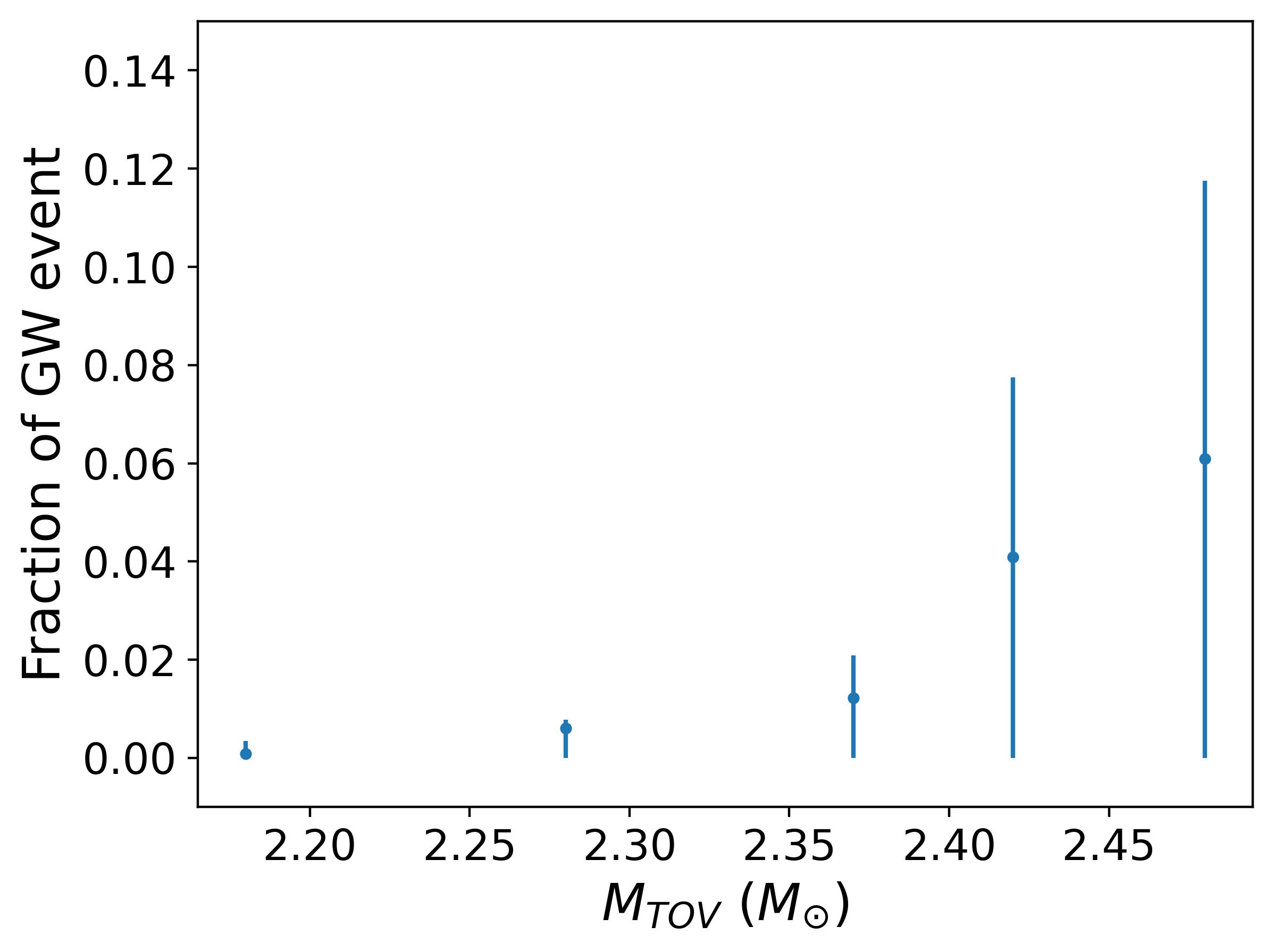}
    \caption{Fraction of detectable millisecond magnetars of the well-localized GW events ($\leq$ 50~deg$^2$) for the different EoS considered, obtained from LVKI O5 simulations and assuming MXT X-ray sensitivity.  
    }
    \label{det_rate}
\end{figure}

To conclude on our results, we also provide the fraction of events among the well-localized GW events ($\leq$ 50~deg$^2$) that would lead to a detection of a millisecond magnetar with MXT. For this, we give the results over all considered EoS, obtained from the LVKI O5 simulations. Figure \ref{det_rate} shows that the fraction of detectable magnetars ranges from 0\% to 6\% depending on the EOS. Considering the highest $M_{\rm TOV}$ EoS and its associated uncertainty, this fraction may rise to $\sim$ 12\% maximum.

\subsection{Observation strategy}
\label{sect:obs_strategy_magnetar}
The GW detection induces a bias on the BNS inclination, in other words on the potential EM observation viewing angle. Moreover, due to the physics of the ejecta, there exists an inherent dependence of the EM luminosity on the observer viewing angle. This has been recovered in our simulation results, as presented in Sect. \ref{sect:freevstrapped_results}, where we reported the repartition of fluxes in the free and trapped zone depending on GW detector configuration.
This will be a major feature to consider in view of a millisecond magnetar detection. In this section we discuss in more details the expected effect of the observer viewing angle, induced by the combined effects of the GW and EM emissions, on the potential detections arising from the synthetic BNS population. This enables us to outline a simulation-informed observation strategy for optimal chances of detection. Throughout this discussion, we consider the millisecond magnetar EM counterpart using the 10$^{15}$G magnetic field results. 
 
The trapped zone is the one furthest away from the jet, centered on the edge-on viewing angle. Thus, the GW emission is of a lesser amplitude in this zone, but it covers the widest solid angle. This implies that, on one hand, the systems in this configuration will be out of reach of GW interferometers at a lower distance than for systems in a free zone configuration. On the other hand, the larger number of systems expected in this configuration implies a significant number of well-localized GW detections in this configuration at lower distances, provided the GW interferometers have the sensitivity to localize them with enough precision.

Concerning the EM emission, the dynamical ejecta, of a high opacity, reprocesses significantly the emission in the trapped zone (see Fig. \ref{obs_geom}). This means a lower luminosity, but also entails an emission that could last longer than the magnetar lifetime, which is particularly interesting for a millisecond magnetar detection prospect, where the time to process the GW alert and to direct instruments in the correct localization is critical. In this case, the use of sensitive X-ray pointed instruments such as MXT and FXT are to be favored. 
 
Since the ejecta is not as dense in the free zone, the luminosity in this viewing angle is higher than in the trapped zone, but typically shorter-lived. For such a brief emission, wide FoV X-ray instruments are naturally considered to observe the EM counterpart, as such instruments enable observers to find a source in large sky regions in a single exposure, sometimes without even requiring a slew (in which case the observations could start from the time of the merger). Thus, provided the systems in this configuration are not too distant, we suggest to approach the millisecond magnetar detection using instruments such as WXT, which have been shown in Sect. \ref{sect:freevstrapped_results} to be suited to detect most magnetars in the free zone (except for ET with CE; see below).

The LVK O4 run maximum BNS range is that of LIGO with 175~Mpc. A significant portion of this distance is also covered by Virgo, thanks to its 50$-$55~Mpc BNS range. Considering these features, for O4, about 75\% of the $\sim$2~yr$^{-1}$ well-localized GW detections are BNS detected in a trapped zone configuration. 
Our results described in Sect. \ref{sect:freevstrapped_results} indicate that from this sample of well-localized GW BNS mergers, the detection of the X-ray EM counterpart also occurs predominantly in the trapped zone configuration, with 75\% of cases, highlighting the importance of more sensitive pointed X-ray instruments for a detection under O4. 
For O5 however, the LIGO range is expected to increase by almost a factor of two compared to O4, while Virgo and KAGRA will not improve their performance as much. Consequently, under the maximum BNS range of the GW fixed by either Virgo or KAGRA, the trapped zone detection of a millisecond magnetar will be favored in the same manner as for O4. For mergers seen with LIGO only, at larger distance, the more important GW amplitude of the free zone will favor the GW detections in this configuration.
As a consequence, the GW results from the $\sim$ 15~yr$^{-1}$ well-localized detections during this observing run are somewhat evenly divided in the free and trapped zones. 
With current GW interferometers, the emission duration is the main limitation for the EM detection, more so than the distance. Thus, 65\% of EM counterpart millisecond magnetar detections for O5 happen in the trapped zone due to their longer duration emission. 
We conclude that for LVK(I) observing runs, the trapped zone, and thus the use of sensitive X-ray pointed instruments, are of particular interest for a millisecond magnetar detection. 

Contrarily to LVK(I) observing runs, for the next generation of GW interferometers, the GW distance reached will be limiting the EM detection, in particular for systems viewed in the trapped zone. 
For ET in a triangle configuration, due to the large distances detectable but the moderate localization ability, 93\% of GW detected systems are expected to be in the free zone, accounting for 80\% of EM counterparts detections.
The ET 2L configuration would increase the number of overall GW detection compared to the triangle configuration, slightly favoring the trapped zone thanks to its ability to better locate in particular close-by BNS systems. 
The localization ability is largely improved by the addition of a CE interferometer to ET in a triangle: the expected rate of well-localized GW events extends from $\sim$ 43~yr$^{-1}$ for ET alone to $\sim$ 21 770~yr$^{-1}$ for ET with CE. With the addition of CE, the number of well-localized GW detections under ET is of 44\% in the free zone, but the distances involved make fluxes fainter, implying that 70\% of EM counterpart detections are happening in the free zone where the emission is typically brighter. Thus, wider FoV X-ray instruments are suited for millisecond magnetar detections under the next generation of GW interferometers, along with more sensitive pointed instruments for particularly distant events, whose X-ray emission has been made fainter and longer duration due to cosmological redshift. An observation strategy involving on jet axis detections could be considered to probe such distant events.

To conclude, no matter the configuration, we emphasize the need for the narrow FoV and wide FoV X-ray instruments for a millisecond magnetar detection, with adjusted observation strategies. Narrow FoV X-ray instruments are mostly suited to observe closer, more reprocessed, longer-lasting emission, i.e typically millisecond magnetars in a trapped zone configuration, which is the configuration favored for current GW interferometers. Less sensitive, wider FoV instruments, are adapted to detect X-rays from distant, short emission duration systems, i.e., millisecond magnetars in a free zone and jet zone\footnote{For a study on joint ET GW and GRB detections, see \cite{Branchesi_2023}.} configurations, and should play particularly an important role under the next generation of GW interferometers, in particular if they gain in sensitivity compared to present-day X-ray instruments.

\section{Discussions} 
\label{sect:simu_discussion}
In this section we further discuss the implications of the results, and the major uncertainties affecting our predictions. We first discuss the uncertainty on the NS mass distribution in Sect. \ref{sect:mass_distrib_discussion}. Then, we review the constraints we can infer from existing observations in Sect. \ref{sect:obs_conditions_discussion}. Finally, we discuss the impact of uncertainties on detection prospects in Sect. \ref{sect:param_degen_discussion}.

\subsection{Uncertainty on the NS mass distribution}
\label{sect:mass_distrib_discussion}

Considering that the mass is the main factor determining the nature of the BNS merger remnant, the NS mass distribution is critical. 
The masses of galactic NS are the most well-constrained. The Galactic double NS systems indicate a Gaussian mass distribution, with the most likely values of the mean and width to be $\mu$ = 1.33~$M_\odot$ and $\sigma$ = 0.09~$M_\odot$ \citep{ns_mass_gal}. However, there is a disagreement between these radio observations of Galactic NS and the two GW events, GW 190425 \citep{gw190425} and GW 170817 \citep{170817}: the masses of the NS components of GW BNSs appear larger than those of Galactic double NSs \citep{bns_not_gal,mass_distrib}. To resolve this observational tension, \cite{Galaudage_2021} and \cite{Farrow_2019} suggest a model involving two independent mass distributions for each binary component. These two mass distributions aim at describing double NS systems, and extragalactic BNS by assuming they originate from the same population. However, various observations tend to indicate that this assumption might be wrong. Indeed, the inferred GW population has a greater prevalence of high-mass NSs.
Namely, for Galactic double NS-based models, the high chirp mass GW 190425 \citep{gw190425} is an outlier, and has to be treated as part of a subpopulation. Additionally, galactic binaries rarely display large mass ratios \citep{equal_mass_ratio_galactic}, contrary to what is suggested by the large amount of ejecta involved in GW 170817 \citep{GW170817_0.1Msol}. One possible explanation for the different mass distributions between Galactic and GW binaries could be that at least some of the high-mass GW events such as GW 190425 may belong to a distinct, fast-merging and unseen in radio subpopulation formed through unstable mass transfer (e.g., \citealt{Galaudage_2021}). Our distribution of mass inferred from LVK O3 observations \citep{mass_distrib} was chosen to best reproduce the GW population despite the uncertainties.

\subsection{Constraints from existing observations}
\label{sect:param_degen_discussion}
In this work, the parameters are degenerate, as multiple unconstrained parameters have an influence on the same observational features. 
Namely, the efficiency of conversion into X-rays $\eta_{\text{dip}}$ and the dipolar magnetic field of the magnetar $B$ both affect the X-ray plateau luminosity. The magnetic field dictates the extraction efficiency of the NS rotational energy, while $\eta_{\text{dip}}$ dictates the conversion efficiency of that energy into X-rays.

The emission of the isolated NS in X-rays is known for Galactic NSs, namely as X-ray pulsars (see also AXPs; e.g., \citealt{axp}). The NS X-ray emission is usually rotation-powered and mediated by the NS magnetic field. However, the detailed physical origin of this nonthermal X-ray emission is uncertain, and depends on a number of parameters (density of the medium, efficiency of the radiation-matter interaction, time since coalescence, etc.). Estimating it for the millisecond magnetar is thus a complex task. 
An efficient and simple way to predict the magnetar X-ray emission is to fix the efficiency parameter $\eta_{\text{dip}}$. The issue is that $\eta_{\text{dip}}$ depends on the NS rotation and magnetic field: these two parameters are high for a millisecond magnetar, while confident observations of NS only fulfill one of the two (i.e., millisecond X-ray pulsars have a lower magnetic field, and magnetars seen through their outburst are slow-rotating). For our study, we have chosen to set $\eta_{\text{dip}}$ conservatively at one of the lowest values proposed in the literature to avoid overestimating fluxes, and to focus the study on exploring the parameters intrinsic to the NS (EoS and magnetic field). 
\cite{efficacite} found a mean value of the spin-down power to X-ray conversion parameter of $\eta_{\text{dip}}$ = 0.01 from the observations of rotation-powered pulsars and their pulsar wind nebulae. Similarly, \cite{sd_pulsar} conducted a study over different types of NS (millisecond, old, Geminga-like, Vela-like, Crab-like, and pulsars), and found a broad range of plausible values, with $\eta_{\text{dip}}$ = 0.01 within the reported range: $\eta_{\text{dip}} =$ 0.001 $-$ 0.80. For a different type of NS, \cite{CDFXT2} fitted observations of a magnetar powered X-ray transient, and argued that reasonable magnetar parameters can be obtained if $\eta_{\text{dip}}$ is of the order of 0.01.
Based on these results, we decided to align with the previous authors, and fix the efficiency of conversion of the magnetar wind injection into X-ray luminosity at $\eta_{\text{dip}}$ = 0.01. With this efficiency parameter that can be considered conservative, a majority of predicted are below or at the limit of MAXI nondetections.

Last, our results are marginally compatible with Swift sGRB observations. According to \cite {rowlinson_2013}, the fraction of Swift sGRBs comprising the plateau and sharp luminosity decay features, i.e., compelling millisecond magnetar central engine candidates, is of 14\% to 28\%. Similarly, \cite{Guglielmi_2024} suggest that the fraction of magnetar central engine lies around 15\% - 26\%, while \cite{magnetar_central_engine_swift} find a fraction of 22\%. The lower fraction of millisecond magnetar in our prediction compared to observations may be an indicator that the EoS is stiff (i.e., close to that of the DDME2 EoS, with $M_{\rm TOV}$ = 2.48 $M_\odot$), a reflection of our conservative hypothesis, or a combination of the two. 
That said, the paradigms that all sGRB have BNS merger for progenitors, or that all BNS mergers lead to GRB and GRB-related emissions, can also be discussed, impacting this statement.
In light of this discussion, we emphasize that our results predict a realistic order of magnitude.

\subsection{Observation prospects}
\label{sect:obs_conditions_discussion}
The prediction of millisecond magnetar detections is also complex because of the uncertainty on instruments characteristics. For O4, these characteristics are more certain; for the GW interferometers and especially for the X-ray instruments performances. However, the delay between the merger and the beginning of X-ray observations is highly uncertain and dependent on many factors. 
Additionally, a detection in O4 is unlikely: the maximum rate predicted by the high $M_{\rm TOV}$ EoS is of $\sim$0.15~yr$^{-1}$, which is consistent with the lack of observations so far. That said, this result was expected, as our chosen BNS merger GW detection rate was adapted to the nondetection of a GW BNS signal at 2/3 of the O4 run. It should be noted that considering the absence of unambiguous and rapid BNS merger detection at the end of O4a \citep{O4a_no_det}, even our adapted BNS rate (predicting 1.7 detection per year for O4) may be considered optimistic, to an extent dependent on the effective BNS ranges of LIGO and Virgo.

The O5 run is foreseen to provide more insight into BNS mergers. Indeed, the detection rate per year of a high $M_{\rm TOV}$ millisecond magnetar with a BNS merger rate of $R = $ 100~Gpc$^{-3}$~yr$^{-1}$ reaches unity for this run. With a more optimistic merger rate of $R = $ 1000~Gpc$^{-3}$~yr$^{-1}$, this detection rate goes up to 10~yr$^{-1}$. Even if a millisecond magnetar is not detected, a BNS merger signal will allow us to refine the merger rate, as well as GW BNS orbital parameters, depending on the precision of the observed signal. In the event of a pessimistic merger rate of $R = $ 10~Gpc$^{-3}$~yr$^{-1}$, despite an unlikely millisecond magnetar detection (0.1~yr$^{-1}$ rate), the observation of at least one BNS merger is very likely under O5, with 1.5~yr$^{-1}$ BNS merger detections within LIGO BNS range. We find that one or even multiple detection(s) of a millisecond magnetar is likely in the era of the next generation of GW interferometers. However, it depends significantly on the final configuration of the interferometers, which is for now highly unclear.  

In parallel, additional indicators of the remnant nature may be provided by GW post-merger signals, expected to be detectable by higher performance designs of the next generation of GW interferometers. First, the GW ringdown signal from the newly formed and oscillating remnant, lying at the high-frequency end of the spectrum ($\geq$ 2 kHz for an NS remnant) may be detected by CE if it features an interferometer of 40~km arms \citep{ringdown_CE}; for ringdown detections prospects with ET see \cite{Branchesi_2023}. Secondly, continuous GW (CW) are an expected emission from a stable (or sufficiently long-lived), rapidly rotating, high magnetic field NS (e.g., \citealt{Sarin2021}; see \citealt{cw170817} for GW170817 long-lived remnant search). This would concern millisecond magnetar formed close-by, i.e., within 20~Mpc for Advanced LIGO and 450~Mpc for the ET (assuming a rather high ellipticity of 10$^{-2}$; see \citealt{cw_distances}). These two post-merger GW signals might be complementary to X-rays to probe the remnant nature over a range of lifetimes, notably along with optical emission, expected to carry signatures of an NS collapsing as early as $\sim 100$ ms after the merger, potentially manifesting as an unusually luminous and blue kilonova (\citealt{Metzger_review, merger_nova}; see also \citealt{ms_mag_det_optical} for detection prospects).

Last, we emphasize that our results evaluated from one hour after the merger might be unrealistic because it assumes either that the localization accuracy of the GW alert is high or that the X-ray instrument is almost immediately repointed after the processing of the GW alert and finds the remnant in one of the first tiles. A beginning of observations at one hour after the merger likely concerns golden-events only. 
That said, our results confirm that the SVOM observation strategy (see Sect. \ref{sect:obs_strategy_magnetar}) is suited for a millisecond magnetar detection. Indeed, a majority of simulated fluxes are above MXT sensitivity threshold for the nominal ToO procedure following a multi-messenger event, and the corresponding exposure time was chosen the shortest so as to cover the GW errorbox as fast as possible.

\section{Summary and conclusions}

We simulated the GW emission and X-ray light-curve from BNS mergers leaving behind a millisecond magnetar remnant. We studied two representative off-axis viewing angles: the trapped zone around the equator with a larger amount of ejecta, and the free zone closer to the jet axis, with fewer ejecta. We implemented polynomial fits from dedicated mergers simulations to compute the opacity and masses of the dynamical and post-merger ejecta. To connect theoretical modeling with observational prospects, we applied these simulations to a synthetic population of BNS systems. We assessed the millisecond magnetar detectability in the near and distant future, considering current GW interferometers (LVK(I)) along with next-generation facilities. The purpose of this work was twofold: first, to state our current and future constraining ability of the nature of the BNS merger remnant, and then, in the case of a millisecond magnetar remnant, to increase our detection chances by suggesting simulation-informed observation strategies. 

We found that approximately 2\% to 16\% of the mergers may form long-lived millisecond magnetars, which are susceptible to collapse into a BH as they spin down. For this subset, we predicted detection rates and explored how intrinsic magnetar parameters affect observability. Our results, which depend on the EoS, suggest that first detections might be made with the current GW network, especially during the O5 run, and they become likely with next-generation interferometers. Intermediate to long term, millisecond magnetar predictions and detections, in particular, for SMNS, might push the constraints on the EoS.

We identified the time of maximum detectability to be about two hours post merger for current GW instruments and roughly three hours for future GW detectors because of cosmological redshift. In the near term, a first detection is most likely in the trapped-zone configuration, requiring pointed X-ray instruments such as SVOM/MXT or EP/FXT. For more distant events, we emphasize the need for synergy between focusing telescopes with a small FOV and wide-field monitors. This role is expected to be fulfilled by THESEUS. These times of maximum detectability were reported for the first time and are particularly interesting for the prospect of detecting a millisecond magnetar, where the time to process the GW alert and to point instruments to the correct localization is critical. We placed moderate constraints on EoS–magnetic field combinations capable of explaining long-duration X-ray emission. In particular, magnetars with field strengths of $10^{16}$~G can account for X-ray plateaus lasting $\gtrsim 30$ minutes only if they are stable, which in turn requires a stiff EoS. Our results indicate, on the other hand, that the best millisecond magnetar detectability occurs for magnetic fields of $\sim 7.5 \times 10^{14}$~G, which should be observable from the start of observations (i.e., one hour after merger), whereas lower-field magnetars ($\sim 10^{14}$~G) would only become detectable hours after the merger. 
These results help us to better understand the emission of the millisecond magnetar and reveal that the detection of its spin-down emission may already be within reach. This motivates further theoretical and observational efforts, as the first millisecond magnetar detection would be a significant advancement in the fields of merger, GRBs, and NS physics.

\begin{acknowledgements}
    The authors thank Ulyana Dupletsa for the guidance on the extraction of ET simulated data, and Francesco Pannarale and David Keitel for comments. ARS, JG, RR, and MB acknowledge support from the European Research Council (MagBURST grant 715368). 
    This project has received funding from the European Union's Horizon Europe research and innovation programme under the Marie Sk\l{}odowska-Curie grant agreement No. 101064953 (GR-PLUTO).
    M.B. acknowledges also the support of the French Agence Nationale de la Recherche (ANR), under grant ANR-24-ERCS-0006 (project BlackJET).

\end{acknowledgements}

\bibliographystyle{aa} 
\bibliography{biblio.bib} 

\newpage

\begin{appendix}
\section{Cosmological effects in the simulation of next generation GW interferometers}
\label{ap:z_effect}

Due to the cosmological distances up to which ET will be able to detect, the effect of the redshift caused by the Universe expansion must be considered. We assume a flat $\Lambda$CDM cosmology, with the Hubble constant $H_0 =$ 67.7 $\rm$ km~s$^{-1}$~Mpc$^{-1}$ and the dark matter fraction $\Omega_M =$ 0.315.

First, the redshift $z$ has a time dilation effect on the arrival of photons, that we accounted for by multiplying time points by a factor $(1+z)$.
Furthermore, the effect of the expansion on the photon energy is crucial to estimate the received flux; for a bolometric flux, it is fully included by using the luminosity distance $D_L$ instead of the classical distance $D$ to compute the flux (as in Eq. \eqref{eq:flux_z}).

To reproduce the observable X-ray flux however, the effect on the spectra also has to be considered. Indeed, the number of received photons in a given energy range is modified by the effect of the redshift, depending on the energy distribution (i.e., the spectra), while the detecting energy band stays the same: the observer detects a different part of the spectra than the one emitted in the source frame. 
The blackbody emission is estimated by integrating the Planck emission law over frequencies shifted by a $(1 + z)$ factor. For the spin-down luminosity, we have to first estimate the X-ray spectra of the emission, which was so far unspecified. 

The spin-down spectrum is assumed to follow a power law of photon index $\Gamma$, that we assume in the magnetar frame, based on GRB X-ray afterglow and X-ray pulsar observations, to be 
\begin{equation}
    N = k E^{-\Gamma},
\end{equation}
with $N$ the number of emitted photons per energy bin (photons cm$^{-2}$ s$^{-1}$ keV$^{-1}$), $k$ a normalization factor, and $E$ the energy in keV. 

Thus finally, the flux received from the magnetar, in the observer frame and for the observable X-ray flux, is expressed as

\begin{equation}
\label{eq:flux_z}
    F_{obs} = \frac{ \mathcal{L} (1+z)^{-\Gamma}}{4 \pi D_L^2}, 
\end{equation}
with $\mathcal{L}$ the luminosity over MXT band in the magnetar frame, and $D_L$ the luminosity distance. Finally, for the numerical application of this formula, $\Gamma$ has to be evaluated. Based on rotation powered pulsar observations by \cite{efficacite} and \cite{sd_pulsar}, and one magnetar candidate observation by \cite{CDFXT2}, we fixed the value of the photon index at $\Gamma = 2$.

\section{Accretion and propeller regimes}
\label{ap:propeller_regime}

To simulate the effects of accretion and propeller regimes on the luminosity, we adopt the formalism used by \cite{Gompertz}. Let us consider a millisecond magnetar surrounded by a disk of accreted matter. The matter forming this disk, “falls back” onto the magnetar after coalescence (known as “fallback accretion”). Then, following interaction with the magnetar's powerful magnetic field, this matter can under certain conditions be accelerated to super-Keplerian speeds. This results in the ejection of matter, as centrifugal force overcomes the magnetar's gravitational attraction. \\

The effect of the magnetic field on the accretion disk, determining this “magnetic propulsion” effect, is quantified by the magnetic pressure exerted on matter at a radius r:
\begin{equation}
    P_{mag} = \frac{\mu^2}{2 \mu_0 r^6}.
\end{equation}

Matter from the accretion disk exerts its own force (ram pressure):
\begin{equation}
    P_{ram} = \frac{\Dot{M}}{8\pi r^2} \left(\frac{2 G M}{r}\right)^{1/2}
\end{equation}

Equating these two formulas gives the typical radius at which matter falling towards the magnetar is strongly influenced by the dipole field, known as the Alfv\'en radius. In other words, the Alfv\'en radius is the radius at which the disk is truncated by the effect of the magnetic field. It is given by

\begin{equation}
    r_m = \mu^{4/7}(GM)^{-1/7}\Dot{M}^{-2/7}.
\end{equation}

Another characteristic radius is the co-rotation radius $r_{corot}$, i.e., the distance from the star where the centrifugal force of a particle co-rotating with the NS surface compensates for gravitational attraction. In other words, the co-rotation radius is the radius where the accretion disk rotates at the same angular velocity as the star. Its expression is as follows:
\begin{equation}
    r_{corot} = \left(\frac{GM}{\Omega^2}\right)^{1/3}.
\end{equation}

By comparing these two characteristic radii, we can determine the emission regime. If $r_{corot} < r_m$, we are in the “propeller” regime: particles in the accretion disk are accelerated to super-Keplerian speeds and ejected. This ejection causes a “propeller” emission, described by the second term of Eq. \ref{L_prop} (in the case of a matter ejection, $\Dot{M} < 0$, the contribution to luminosity of this term is then positive). If, on the other hand, $r_{corot} > r_m$, the inner matter is slowed down, eventually falling onto the central object and accelerating its rotation. This corresponds to an accretion regime, quantified by the accretion torque $N_{acc}$ (positive). The influence of this scenario can be seen in the two terms of the “propeller” luminosity expression. Accelerating angular velocity reduces the rotation $\Omega$, giving the first term, whose contribution is always negative. For the second term, this time with matter accretion $\Dot{M} > 0$, giving a second negative contribution to luminosity. The accretion regime thus always lowers the overall luminosity; however, the acceleration of rotation tends to increase the dipole luminosity accordingly (see Eq. \ref{L_dip}).

\begin{equation}
    L_{prop} = -N_{acc} \Omega - \frac{GM \Dot{M}}{r_m}.
    \label{L_prop}
\end{equation}

Finally, the magnetar X-ray luminosity in the free zone is expressed as
\begin{equation}
   L^X_{\text{free}} = e^{-\tau} (\eta_{\text{dip}} L_{\text{dip}} + \eta_{\text{prop}}L_{\text{prop}}),
\end{equation}
with $\eta_{\text{prop}}$ the conversion efficiency
of kinetic energy to EM radiation for propelled material, set at 0.4 in our simulations based on the highest estimations of \cite{Gompertz}.

\section{Flux evolution with time}
\label{ap:flux_vs_time}

\begin{figure*}
    \centering
    \includegraphics[width=.93\linewidth]{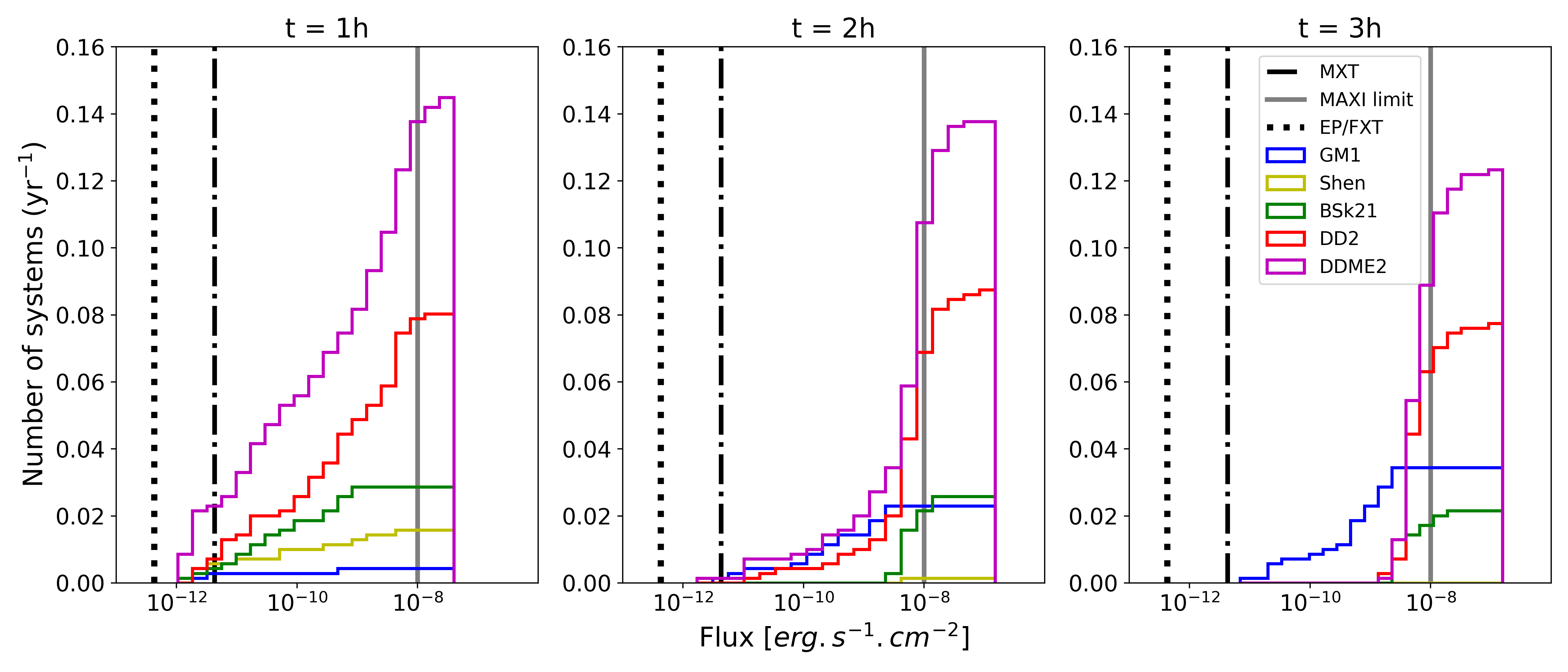}
    \caption{Same as Fig. \ref{fluxvstime_ETtriangle} but for O4. 
    }
    \label{fluxvstime_O4}
\end{figure*}

\begin{figure*}
    \centering
    \includegraphics[width=.93\linewidth]{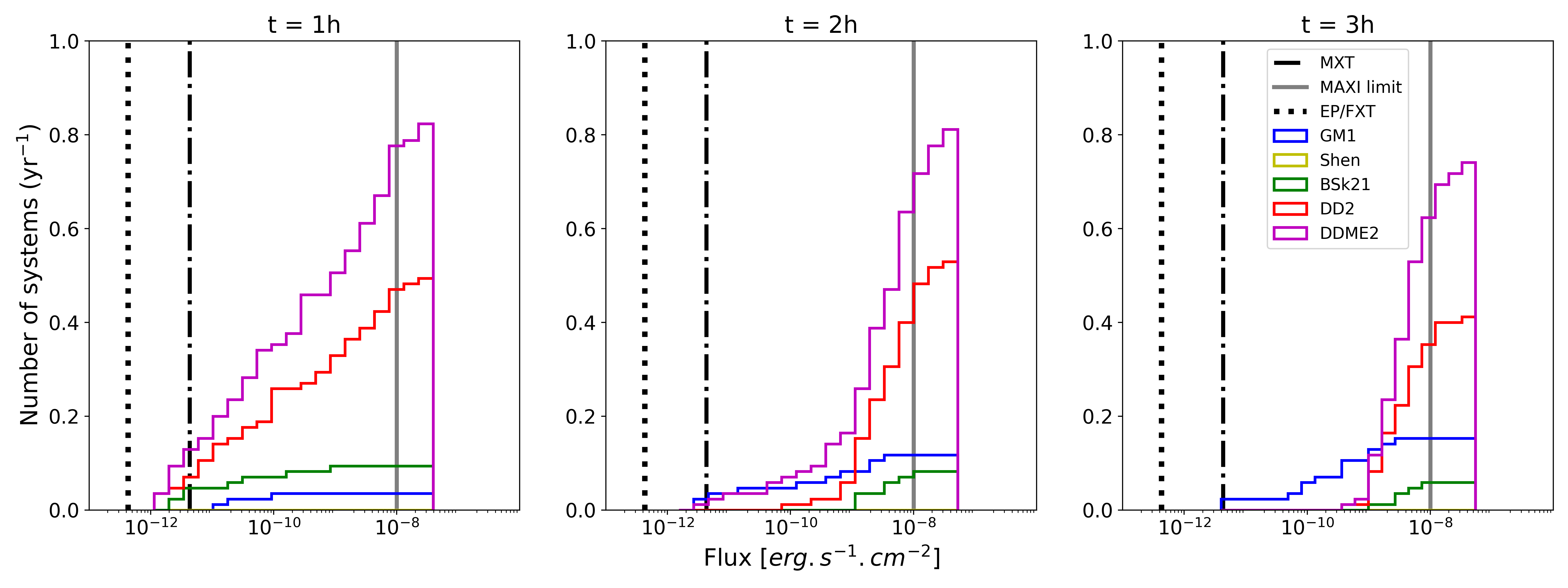}
    \caption{Same as Fig. \ref{fluxvstime_ETtriangle} but for O5.}
     \label{fluxvstime_O5}
\end{figure*}

\begin{figure*}
    \centering
    \includegraphics[width=.93\linewidth]{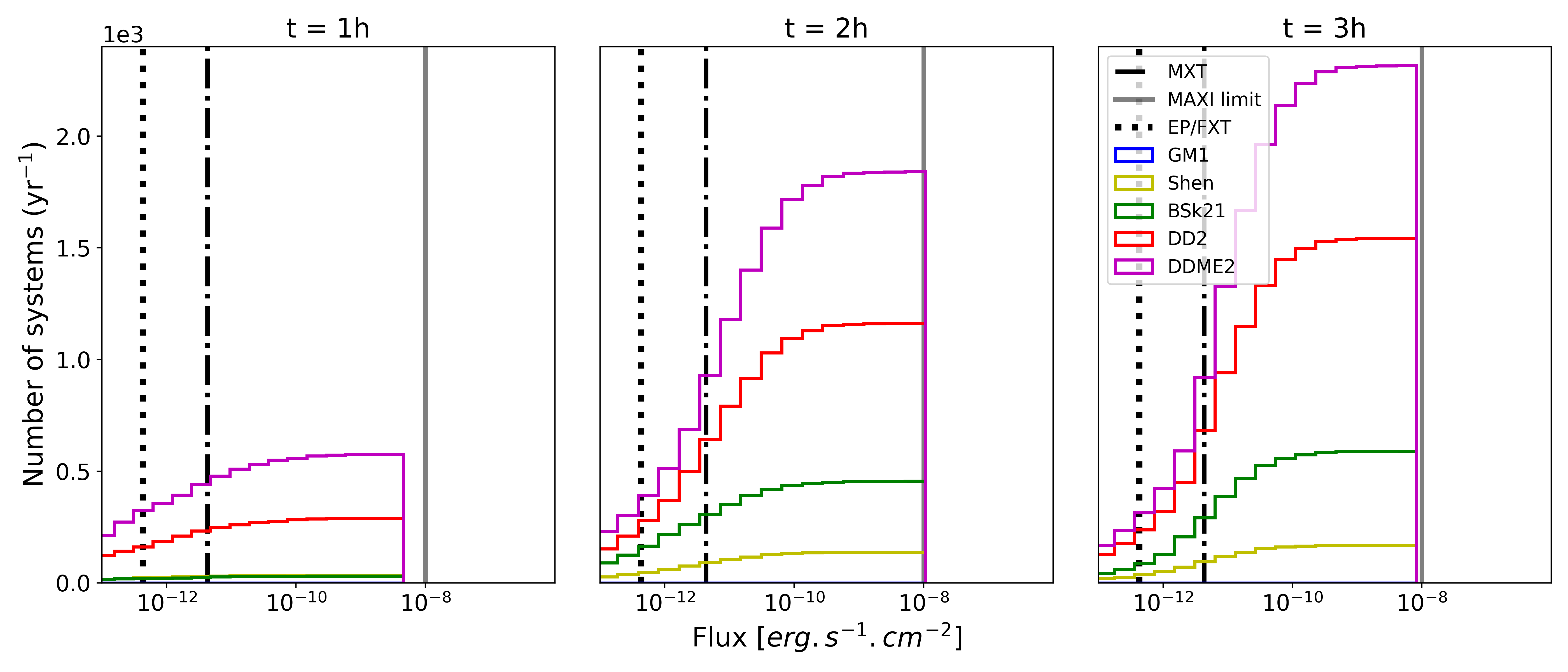}
    \caption{Same as Fig. \ref{fluxvstime_ETtriangle} but for ET $\Delta$ 10~km + 1 CE 40~km.}
    \label{fluxvstime_ET_CE}
\end{figure*}

The detailed evolution of the millisecond magnetar flux with time is reported in Fig.~\ref{fluxvstime_O4} for O4 and in Fig. \ref{fluxvstime_O5} for O5. For the next generation of GW interferometers, the same figure is provided in Fig. \ref{fluxvstime_ETtriangle} for ET alone in a triangle configuration, and in Fig. \ref{fluxvstime_ET_CE} for ET along with CE.

\label{LastPage}

\end{appendix}

\end{document}